\newif\ifTR
\TRtrue
\documentclass[a4paper,USenglish]{lipics-v2016}

\usepackage{microtype}
\usepackage{etex}
\usepackage{amsmath}
\usepackage{amsthm}
\usepackage{amssymb}
\usepackage{stmaryrd}
\usepackage{xcolor}
\usepackage{xspace}
\usepackage{mathpartir}
\usepackage{tikz}

\usepackage{dsfont}
\usepackage{oz}
\renewcommand{\MathparLineskip}{\lineskiplimit=3pt\lineskip=3pt}

\newtheorem{proposition}[theorem]{\bfseries\sffamily Proposition}
\newenvironment{myproof}[1][Proof]{$\blacktriangleright$\;{\textsf{#1}.\;}}{\hfill$\square$}

\ifTR
\title{A Generic Approach to Flow-Sensitive Polymorphic Effects (Extended Version)}
\author{Colin S.\ Gordon}
\affil{Department of Computer Science, Drexel University, Philadelphia, PA, USA\\\texttt{csgordon@drexel.edu}}
\authorrunning{C.\ S.\ Gordon} \Copyright{Colin S.\ Gordon}\else
\title{A Generic Approach to Flow-Sensitive Polymorphic Effects}
\author{Colin S.\ Gordon}
\affil{Drexel University\\\texttt{csgordon@drexel.edu}}
\authorrunning{C.\ S.\ Gordon} \Copyright{Colin S.\ Gordon}\fi

\DOIPrefix{}

\ifTR
\EventEditors{}
\EventNoEds{0}
\EventLongTitle{Extended version of a paper to appear at ECOOP 2017}
\EventShortTitle{}
\EventAcronym{}
\EventYear{2017}
\EventDate{}
\EventLocation{}
\EventLogo{}
\SeriesVolume{}
\ArticleNo{}
\else
\EventEditors{Peter M\"uller}
\EventNoEds{1}
\EventLongTitle{31st European Conference on Object-Oriented Programming
(ECOOP 2017)}
\EventShortTitle{ECOOP 2017}
\EventAcronym{ECOOP}
\EventYear{2017}
\EventDate{June 18--23, 2017}
\EventLocation{Barcelona, Spain}
\EventLogo{}
\SeriesVolume{74}
\ArticleNo{20}
\fi

\begin{document}
\maketitle
\begin{abstract}
Effect systems are lightweight extensions to type systems that can verify a wide range of important properties with modest developer burden.  But our general understanding of effect systems is limited primarily to systems where the order of effects is irrelevant.
Understanding such systems in terms of a lattice of effects grounds understanding of the essential issues, and provides guidance when designing new effect systems.
By contrast, sequential effect systems --- where the order of effects is important --- lack a clear algebraic characterization.

We derive an algebraic characterization from the shape of prior concrete sequential effect systems.
We present an abstract polymorphic effect system with singleton effects parameterized by an effect quantale --- an algebraic structure with well-defined properties that can model a range of existing order-sensitive effect systems.  We define effect quantales, derive useful properties, and show how they cleanly model a variety of known sequential effect systems.
We show that effect quantales provide a free, general notion of iterating a sequential effect, and that for systems we consider the derived iteration agrees with the manually designed iteration operators in prior work.
Identifying and applying the right algebraic structure led us to subtle insights into the design of order-sensitive effect systems, which provides guidance on non-obvious points of designing order-sensitive effect systems.
Effect quantales have clear relationships to the recent category theoretic work on order-sensitive effect systems, but are explained without recourse to category theory.  In addition, our derived iteration construct should generalize to these semantic structures, addressing limitations of that work.
\end{abstract}

\section{Introduction}
Effect systems are a well-known lightweight extension to standard type systems, which are capable of verifying an array of useful program properties with modest developer effort.
They have proven useful for enforcing error handling~\cite{vanDooren2005,benton2007exceptions,gosling2014java},
ensuring a variety of safety properties for concurrent programs~\cite{safelocking99,rccjava00,objtyrace99,boyapati01,boyapati02,flanagan2003atomicity},
purity~\cite{Hunt2007sealing,Fahndrich2006language},
safe arena-based memory management~\cite{lucassen88,talpin1992polymorphic,Tofte1994Regions}, and more.
Effect systems extend type systems to track not only the shape of and constraints on data, but also a summary of the side effects caused by an expression's evaluation.
Java's checked exceptions are the best-known example of an effect system --- the effect of an expression is the set of (checked) exceptions it may throw --- and other effects have a similar flavor, like the set of heap regions accessed by parallel code, or the set of locks that must be held to run an expression without data races.

However, our understanding of effect systems is concentrated in the space of systems like Java's checked exceptions, where the \emph{order of effects} is irrelevant: the system does not care that an \texttt{IllegalArgumentException} would be thrown before any possible \texttt{IOException}.
Effects in such systems are characterized by a join semilattice, which captures exactly systems where ordering is irrelevant (since the join operation is commutative and associative).
This is an impressively large and useful class of systems, but the assumption that order is irrelevant leaves some of the more sophisticated effect systems for checking more powerful properties out of reach.  We refer to this class of effect systems --- the traditional default --- as \emph{commutative} effect systems, to contrast against the class we study in this paper.  The alternative class of effect systems, where the order in which effects occur matters --- \emph{sequential} effect systems, following Tate's terminology~\cite{tate13}\footnote{These effect systems have been alternately referred to as flow-sensitive~\cite{marino09}, as they are often formalized using flow-sensitive type judgments (with pre- and post-effect) rather than effects in the traditional sense.  However, this term suggests a greater degree of path sensitivity and awareness of branch conditions than most such systems have.  We use Tate's terminology as it avoids technical quibbles.} --- reason directly about the proper ordering of program events.  Examples include
non-block-structured reasoning about synchronization for data races and
deadlock freedom~\cite{boyapati02,tldi12,suenaga2008type},
atomicity~\cite{flanagan2003tldi,flanagan2003atomicity},
and memory management~\cite{crary1999typed}.

Effect system design for the traditional commutative effect systems has been greatly aided in both theory and practice by the recognition that effects in such systems form a bounded join semilattice --- a lattice with top, bottom, and all binary joins (least-upper-bounds).  On the theory side, this permits general formulations of effect systems to study common properties~\cite{marino09,rytz12,BanadosSchwerter2014gradual}.  On the practical side, this guides the design and implementation of working effect systems.  If an effect system is not a join semilattice, why not? (Usually this indicates a mistake.)  Effect system frameworks can be implemented generically with respect to an effect lattice~\cite{rytz12,toro2015customizable}, and in the common case where effects are viewed as sets of required capabilities, simply specifying the capabilities and exploiting the default powerset lattice makes core design choices straightforward.  In the research literature, the ubiquity of lattice-based (commutative) effect systems simplifies explanations and presentations.

Sequential effect systems so far have no such established common basis in terms of an algebraic structure to guide design, implementation, and comparison, making all of these tasks more difficult.
Recent work on semantic approaches to modeling sequential effect systems~\cite{tate13,katsumata14,mycroft16} has produced very general characterizations of the mathematics behind key \emph{necessary} constructs (namely, sequencing effects), but with one recent exception~\cite{mycroft16} does not produce a description that is \emph{sufficient} to model a complete sequential effect system for a real language.
Partly this stems from the fact that the accounts of such work proceed primarily by generalizing categorical structures used to model sequential computation, rather than implementing complete source-level effect systems.  None of this work has directly considered effect polymorphism (essential for any real use), singleton effects (required for prominent effect systems both commutative and sequential), or iteration constructs.  So there is currently a gap between this powerful semantic work, and understanding real sequential effect systems in a systematic way.

We generalize directly from real source-level type-and-effect systems to produce an algebraic characterization for sequential effect systems, suitable for modeling some well-known sequential type-and-effect disciplines, and (we hope) useful for guiding the design of future sequential effect systems.  We give important derived constructions (products, and inducing an iteration operation on effects), and put them to use with an explicit translation between Flanagan and Qadeer's early atomicity type system~\cite{flanagan2003tldi} and a (sequential) equivalent built as an instantiation of our generic sequential type-and-effect system.

Overall, our contributions include:
\begin{itemize}
\item A new algebraic characterization of sequential effect systems --- effect quantales --- that is consistent with existing semantic notions and easily subsumes commutative effects
\item A syntactic motivation for effect quantales by generalizing from concrete, full-featured sequential effect systems. As a result, we are the first to investigate interplay between singleton effects and sequential effect systems in the abstract (not yet addressed by semantic work).  This reveals subtlety in the metatheory of sequential effects that depend on program values.
\item Demonstration that effect quantales are not only general, but also sufficient to modularly define the structure of existing non-trivial effect systems.
\item A general characterization of effect iteration for any sequential effect system given by an effect quantale, including demonstration that the resulting iteration for prior systems (as effect quantales) exactly matches the hand-constructed iteration of the original work.  The form is general enough that it should adapt to semantic characterizations as well.
\item The first generic \emph{sequential} effect system with effect polymorphism.
\item Precise characterization of the relationship between effect quantales and related notions, ultimately connecting the syntax of established effect systems to semantic work, closing a gap in our understanding.
\end{itemize}

\section{Background on Commutative and Sequential Effect Systems}
\label{sec:bg}
Here we derive the basic form of a new algebraic characterization of sequential effects based on generalizing from the use of effects in current sequential effect systems.  The details of this form are given in Section \ref{sec:quantales}, with a corresponding generic type-and-effect system in Section \ref{sec:soundness}.  We refer to the two together as a framework for sequential effect systems.

By now, the standard mechanisms of commutative effect systems --- what is typically meant by the phrase ``type-and-effect system'' --- are well understood.  The type judgement $\Gamma\vdash e : \tau$ of a language is augmented with a component $\chi$ describing the overall effect of the term at hand: $\Gamma\vdash e : \tau \mid \chi$.  Type rules for composite expressions, such as forming a pair, join the effects of the child expressions by taking the least upper bound of those effects (with respect to the effect lattice).  And the final essential adjustment is to handle the \emph{latent effect} of a function --- the effect of the function body, which is deferred until the function is invoked.  Function types are extended to include this latent effect, and this latent effect is included in the effect of function application.  Allocating a closure itself has no meaningful effect, and is typically given the bottom effect in the semilattice:
\begin{mathpar}
\inferrule*[left=T-Fun]{\Gamma,x:\tau\vdash e : \tau' \mid \chi}{\Gamma\vdash(\lambda x\ldotp e) : \tau\overset{\chi}{\rightarrow}\tau' \mid \bot}
\and
\inferrule*[left=T-Call]{
    \Gamma\vdash e_1 : \tau\overset{\chi}{\rightarrow}\tau' \mid \chi_1\\
    \Gamma\vdash e_2 : \tau \mid \chi_2
}{
    \Gamma\vdash e_1\;e_2 : \tau' \mid \chi_1\sqcup\chi_2\sqcup\chi
}
\end{mathpar}
Consider the interpretation for concrete effect systems.  Java's checked exceptions are an effect system~\cite{gosling2014java,vanDooren2005}: the effects are sets of (checked) exceptions ordered by inclusion, with set union as the semilattice join.  The \texttt{throws} clause of a method states its \emph{latent} effect --- the effect of actually \emph{executing} the method (roughly $\chi$ in \textsc{T-Fun} above).  The exceptions thrown by a composite expression such as invoking a method is the union of the exceptions thrown by subexpressions (e.g., the receiver object expression and method arguments) and the latent effect of the code being executed (as in \textsc{T-Call} above).  Most effect systems for treating data race freedom (for block-structured synchronization like Java's \texttt{synchronized} blocks, such as \textsc{RCC/Java}~\cite{rccjava00,Abadi2006}) use sets of locks as effects, where an expression's effect is the set of locks guarding data that may be accessed by that expression.  The latent effect there is the set of locks a method requires to be held by its call-site.  Other effect systems follow similar structure: a binary yes/no effects of whether or not code performs a sensitive class of action like allocating memory in an interrupt handler~\cite{Hunt2007sealing,Hunt2007singularity,Fahndrich2006language} or accessing user interface elements~\cite{ecoop13}; tracking the sets of memory regions read, written, or allocated into for safe memory deallocation~\cite{talpin1992polymorphic,Tofte1994Regions} or parallelizing code safely~\cite{lucassen88,gifford86} or even deterministically~\cite{bocchino09,kawaguchi12}.

But these and many other examples do not care about ordering.  Java does not care which exception might be thrown first.  Race freedom effect systems for block-structured locking do not care about the order of object access within a \texttt{synchronized} block.  Effect systems for region-based memory management do not care about the order in which regions are accessed, or the order of operations within a region.  Because the order of combining effects in these systems is irrelevant, we refer to this style of effect system as \emph{commutative} effect systems, though due to their prevalence and the fact that they arose first historically, this is the class of systems typically meant by general references to ``effect systems.''

Sequential effect systems tend to have slightly different proof theory.  Many of the same issues arise (latent effects, etc.) but the desire to enforce a sensible \emph{ordering} among expressions leads to slightly richer type judgments.  Often they take the form $\Gamma;\Delta\vdash e : \tau \mid \chi \dashv \Delta'$.  Here the $\Delta$ and $\Delta'$ are some kind of pre- and post-state information --- for example, the sets of locks held before and after executing $e$~\cite{suenaga2008type}, or abstractions of heap shape before and after $e$'s execution~\cite{tldi12}.  $\chi$ as before is an element of some lattice.  Some sequential effect systems have both of these features, and some only one or the other.  (These components never affect the type of variables, and stricly reflect some property of the \emph{computation} performed by $e$, making them part of the effect.)
The judgements for something like a variant of Flanagan and Qadeer's atomicity type system that tracks lock sets flow-senstiviely rather than using synchronized blocks
or for an effect system that tracks partial heap shapes before and after updates~\cite{tldi12} might look like the following, using $\Delta$ or $\Upsilon$ to track locks held, and tracking atomicities with $\chi$:
\renewcommand{\MathparLineskip}{\lineskiplimit=3pt\lineskip=3pt}
\begin{mathpar}
    \inferrule{\Gamma,x:\tau;\Upsilon\vdash e : \tau' \mid \chi\dashv\Upsilon'}{\Gamma;\Delta\vdash(\lambda x\ldotp e) : \tau{\xrightarrow{\Upsilon,\chi,\Upsilon'}}\tau' \mid \bot \dashv \Delta}
\and
\inferrule{
    \Gamma;\Delta\vdash e_1 : \tau{\xrightarrow{\Delta'',\chi,\Delta'''}}\tau' \mid \chi_1\dashv\Delta'\\\\
    \Gamma;\Delta'\vdash e_2 : \tau \mid \chi_2\dashv\Delta''
}{
    \Gamma\Delta\vdash e_1\;e_2 : \tau' \mid \chi_1;\chi_2;\chi \dashv\Delta'''
}
\end{mathpar}
The sensitivity to evaluation order is reflected in the threading of $\Delta$s through the type rule for application, as well as through the switch to the sequencing composition $;$ of the basic effects.
Confusingly, while $\chi$ continues to be referred to as the effect of this judgment, the real effect is actually a combination of $\chi$, $\Delta$, and $\Delta'$ in the judgment form.  This distribution of the ``stateful'' aspects of the effect through a separate part of the judgment obscures that this judgment really tracks a product of \emph{two} effects --- one concerned with the self-contained $\chi$, and the other a form of effect indexed by pre- and post-computation information.

Rewriting these traditional sequential effect judgements in a form closer to the commutative form reveals some subtleties of sequential effect systems:
\begin{mathpar}
    \inferrule{\Gamma,x:\tau\vdash e : \tau' \mid (\Upsilon\leadsto\Upsilon')\otimes\chi
}{
    \Gamma\vdash(\lambda x\ldotp e) : \tau{\xrightarrow{(\Upsilon\leadsto\Upsilon')\otimes\chi}}\tau' \mid (\Delta\leadsto\Delta)\otimes\bot
}
\and
\inferrule{
    \Gamma\vdash e_1 : \tau{\xrightarrow{(\Delta''\leadsto\Delta''')\otimes\chi}}\tau' \mid (\Delta\leadsto\Delta')\otimes\chi_1\\
    \Gamma\vdash e_2 : \tau \mid (\Delta'\leadsto\Delta'')\otimes\chi_2
}{
    \Gamma\vdash e_1\;e_2 : \tau' \mid ((\Delta\leadsto\Delta');(\Delta'\leadsto\Delta'');(\Delta''\leadsto\Delta'''))\otimes(\chi_1;\chi_2;\chi)
}
\end{mathpar}
One change that stands out is that the effect of allocating a closure is not simply the bottom effect (or product of bottom effects) in some lattice.  No sensible lattice of pre/post-state pairs has equal pairs as its bottom.  However, it makes sense that some such equal pair acts as the left and right identity for \emph{sequential} composition of these ``stateful'' effects.  In commutative effect systems, sequential composition is actually least-upper-bound, for which the identity element happens to be $\bot$.
We account for this in our framework.

We also assumed, in rewriting these rules, that it was sensible to run two effect systems ``in parallel'' in the same type judgment, essentially by building a product of two effect systems.  Some sequential effect systems are in fact built this way, as two ``parallel'' systems (e.g., one for tracking locks, one for tracking atomicities, one for tracking heap shapes, etc.) that together ensure the desired properties.  The general framework we propose supports a straightforward product construction.

Another implicit assumption in the refactoring above is that the effect tracking that is typically done via flow-sensitive type judgments is equivalent to \emph{some} algebraic treatment of effects akin to how $\chi$s are managed above.  While it is clear we would \emph{want} a clean algebraic characterization of such effects, the existence of such an algebra that is adequate for modeling known sequential effect systems for non-trivial languages is not obvious.  Our proposed algebraic structures (Section \ref{sec:quantales}) are adequate to model such effects (Section \ref{sec:modeling}

Examining the sequential variant of other rules reveals more subtleties of sequential effect system design.
For example, effect joins are still required in sequential systems:
\begin{mathpar}
\inferrule{
    \Gamma\vdash e : \mathcal{B} \mid \chi\quad
    \Gamma\vdash e_1 : \tau \mid \chi_1\quad
    \Gamma\vdash e_2 : \tau \mid \chi_2
}{
    \Gamma\vdash\mathsf{if}\;e\;e_1\;e_2 : \tau \mid \chi\sqcup\chi_1\sqcup\chi_2
}
{\Rightarrow}
\inferrule{
    \Gamma\vdash e : \mathcal{B} \mid \chi\quad
    \Gamma\vdash e_1 : \tau \mid \chi_1\quad
    \Gamma\vdash e_2 : \tau \mid \chi_2
}{
    \Gamma\vdash\mathsf{if}\;e\;e_1\;e_2 : \tau \mid \chi;(\chi_1\sqcup\chi_2)
}
\end{mathpar}
Nesting conditionals can quickly produce an effect that becomes a mass of alternating effect sequencing and join operations.  For a monomorphic effect system, concrete effects can always be plugged in and comparisons made.  However, for a polymorphic effect system, it is highly desirable to have a sensible way to simplify such effect expressions --- particularly for highly polymorphic code --- to avoid embedding the full structure of code in the effect.  Our proposal codifies natural rules for such simplifications.

\section{Effect Quantales}
\label{sec:quantales}
\ifTR
Quantales~\cite{mulvey1986,mulvey1992quantisation} are an algebraic structure originally proposed to generalize some concepts in topology to the non-commutative case.  They later found use in models for non-commutative linear logic~\cite{yetter1990quantales} and reasoning about observations of computational processes~\cite{abramsky1993quantales}, among other uses. Abramsky and Vickers give a thorough historical account~\cite{abramsky1993quantales}.  They are almost exactly the structure we require to model a sequential effect system.  Here we give the original definition, and then relax it slightly to better suit the needs of sequential effect systems.
We establish one very useful property of effect quantales, and show how they subsume commutative effect systems.
We defer more involved examples to Section \ref{sec:modeling}.

\begin{definition}[Quantale~\cite{mulvey1986,mulvey1992quantisation}]
A \emph{quantale} $Q=(E,\wedge,\vee,\cdot)$ is a complete lattice $(E,\wedge,\vee)$ with arbitrary joins and an associative product $\cdot$ that distributes on both sides over arbitrary (including infinite) joins:\\
\centerline{
$a\cdot \left(\bigvee b_i\right) = \bigvee (a\cdot b_i)$
and
$\left(\bigvee b_i\right)\cdot a = \bigvee (b_i\cdot a)$
}
Additionally, a quantale is called \emph{unital} if it includes an element $I$ that acts as left and right unit (identity) for the product --- $I\cdot a=a=a\cdot I$, or in other words $(E,\cdot,I)$ is a monoid.
\end{definition}
Because of the similarity to rings, the join is often referred to as the additive element, while the semigroup or monoid operation is typically referred to as the multiplicative operation.  Because the lattice is complete, it is bounded, and therefore contains both a greatest and least element.

A unital quantale is close to what we require, but just slightly too strong.  In particular, it requires a complete, and therefore bounded lattice, which therefore has a least element.
The \emph{complete} lattice structure with distributive laws make all quantales residuated lattices~\cite[\textsection 2.2, 3.1.3, 3.2, 3.4.14]{galatos2007residuated}, where the bottom element is always nilpotent for multiplication --- $\bot\cdot x = \bot = x\cdot\bot$, for all $x$.
This conflicts with the common practice in prior sequential effect systems of using the bottom element as the identity for composition.
There are also systems where there is no natural bottom element, such as the lockset example we develop later.
Thus we need a slightly more general structure.  In particular, we do not require a bottom element, nor do we have use for a meet operation.  The need to join over empty or infinite sets is also not required by any effect system we know of.  Thus we replace the complete lattice of a standard quantale with a join semilattice, in addition to requiring the unit.
For clarity, we also switch the the suggestive (directional) $\rhd$ for sequencing, rather than the multiplication symbol $\cdot$ or the common practice in work on quantales and related structures~\cite{yetter1990quantales,abramsky1993quantales,galatos2007residuated,kozen1997kleene,pratt1990action} of eliding the multiplicative operator entirely and writing ``strings'' $abc$ for $a\cdot b\cdot c$.
\fi
\begin{definition}[Effect Quantale]
An \emph{effect quantale} $Q=(E,\sqcup,\rhd,\top,I)$ is a join semilattice $(E,\sqcup)$ with top element $\top$ with monoid $(E,\rhd,I)$, such that $\rhd$ distributes over joins in both directions ---
$a\rhd(b\sqcup c)=(a\rhd b)\sqcup(a\rhd c)$ and
$(a\sqcup b)\rhd c = (a\rhd c)\sqcup(b\rhd c)$ --- and $\top$ is a nilpotent element for the monoid ($a\rhd\top=\top=\top\rhd a$).
\end{definition}
\ifTR
An effect quantale is essentially an \emph{upper} unital quantale with nilpotent top.  Other descriptions are apt, as with other composite algebraic structures, and we could have equivalently characterized it as a bounded-join-semilattice-ordered monoid with nilpotent top.  For brevity, we will simply use ``effect quantale.''
\fi

As is standard in lattice theory, we induce the partial order $x\sqsubseteq y\overset{\mathsf{def}}{=} x\sqcup y = y$ from the join operation, which ensures the properties required of a partial order.

We will use the semilattice to model the standard effect hierarchy, using the partial order for subeffecting.  The (non-commutative) monoid operation $\rhd$ will act as the sequential composition.  The properties of the semilattice and distibutivity of the product over joins will permit us to move common prefixes or suffixes of effect sequences into or out of least-upper-bounds of effects, permitting more concise specifications.
Intuitively, the unit $I$ is an ``empty'' effect, which need not be a bottom element.
$\top$ is an error (invalid effect, allowing us to reason about ``undefined'' effect sequences).

\label{sec:quantale-props}

Effect quantales inherit a rich equational theory of semilattices, monoids, and extensive study of ordered algebraic systems~\cite{birkhoff,galatos2007residuated,blyth2006lattices,fuchs2011partially} from their several substructures, providing many ready-to-use properties for simplifying complex effects, and giving rise to other properties more interesting to our needs.
One such example is an important and expected form of monotonicity property: that sequential composition respects the partial order on effects.  In lattice-ordered monoids, this property is called isotonicity, and its proof for complete lattices~\cite[ch.~14.4]{birkhoff} carries over directly to effect quantales because it requires only binary joins:
\begin{proposition}[Isotonicity]
In an effect quantale $Q$, $a\sqsubseteq b$ and $c\sqsubseteq d$ implies that $a\rhd c \sqsubseteq b\rhd d$.
\end{proposition}
\begin{myproof}
Because
$b\rhd d = b\rhd(c\sqcup d) = (b\rhd c) \sqcup (b\rhd d)$, we know
$b\rhd c \sqsubseteq b\rhd d$ by the definition of $\sqsubseteq$. Repeating the reasoning:
\mbox{$b\rhd c = (a\sqcup b)\rhd c = (a\rhd c) \sqcup (b\rhd c)$,}
so $a\rhd c \sqsubseteq b\rhd c$.  The partial order is transitive, thus
$a\rhd c \sqsubseteq b\rhd d$
\end{myproof}

An important litmus test for a general model of sequential effects is that it should subsume commutative effects (modeled as a join semilattice).
This not only implies consistency of effect quantales with traditional effect systems, but ensures implementation frameworks for sequential effects (based on effect quantales) would be adequate for implementing commutative systems as well.
\begin{lemma}[Subsumption of Commutative Effects]
\label{lem:subsume}
Every commutative effect system modeled as a bounded join semilattice yields an effect quantale, such that ordering of individual effects is irrelevant, by using join for the monoid operation.
\end{lemma}
\begin{myproof}
Assume a bounded join semilattice $L=(E,\vee,\top,\bot)$ of effects.  Define a new effect quantale $Q$ as $(E,\vee,\vee,\top,\bot)$ (i.e., reuse the join for the monoid).  $Q$ satisfies the distributivity requirements of the effect quantale definition, and naturally has $\bot$ as the monoid unit.
\end{myproof}

\section{Modeling Prior Sequential Effect Systems with Effect Quantales}
\label{sec:modeling}
Many of the axioms of effect quantales are not particularly surprising given prior work on
sequential effect systems; one of this paper's contributions is recognizing that these axioms are
sufficiently general to capture many prior instances of sequential effect systems. We show two
prominent examples here in detail, and cite further examples.

\subsection{Locking with Effect Quantales}
A common class of effect systems is those reasoning about synchronization --- which locks are held at various points in the program.  In most systems this is done using scoped synchronization constructs, for which a bounded join semilattice is adequate.  Here, we give an effect quantale for flow-sensitive tracking of lock sets including recursive acquisition.
The main idea is to use a multiset of locks (modeled by $\mathcal{M}(S)=S\rightarrow\mathbb{N}$, where the multiplicity of a lock is the number of claims a thread has to holding the lock --- the number of times it has acquired said lock) for the locks held before and after each expression.  We use $\emptyset$ to denote the empty multiset (where all multiplicities are 0).  We use join on multisets to produce least upper bounds on multiplicities, union to perform addition of multiplicities, and set difference for zero-limited subtraction.

\begin{definition}[Synchronization Effect Quantale $\mathcal{L}$]
An effect quantale $\mathcal{L}$ for lock-based synchronization with explicit mutex acquire and release primitives is given by:
\begin{itemize}
\item $E=\mathcal{M}(L)\times\mathcal{M}(L)\uplus\mathsf{Err}$ for a set $L$ of possible locks.
\item $(a,a')\sqcup (b,b') = (a\vee a',b\vee b')$ when both effects acquire and release the same set of locks the same number of times: $b/b'=a/a'$ and $b'/b=a'/a$.  Otherwise, the join is \textsf{Err}.
\item $(a,a')\rhd(b,b')$ is $(c,c')$ for the least $c$ and $c'$ where $a\subseteq c$, $b\subseteq ((c/(a/a'))\cup(a'/a))$ ($b$'s holdings are contained in $c$ less lock releases from the first action, plus the lock acquisitions from the first action), and $c'=(((c/(a/a'))\cup(a'/a))/(b/b'))\cup(b'/b)$ when such a pair exists, and \textsf{Err} otherwise. 
\item $\top=\mathsf{Err}$
\item $I=(\emptyset,\emptyset)$
\end{itemize}
\end{definition}
Intuitively, the pair represents the sets of lock claims before and after some action, which models lock acquisition and release.
$\sqcup$ intuitively requires each ``alternative'' to acquire/release the same locks, while the set of locks held for the duration may vary (and the result assumes enough locks are held on entry --- enough times each --- to validate either element).
This can be intuitively justified by noticing that most effect systems for synchronization require, for example, that each branch of a conditional may access different memory locations, but reject cases where one branch changes the set of locks held while the other does not (otherwise the lock set tracked ``after'' the conditional will be inaccurate for one branch, regardless of other choices).
Sequencing two lock actions, roughly, pushes the locks required by the second action to the precondition of the compound action (unless such locks were released by the first action, i.e.\ in $a/a'$), and pushes locks held after the first action through the second --- roughly a form of bidirectional framing. 

With this scheme, lock acquisition for some lock $\ell$ would have (at least) effect $(\emptyset,\{\ell\})$, indicating that it requires no locks to execute safely, and terminates holding lock $\ell$.  A release of $\ell$ would have swapped components --- $(\{\ell\},\emptyset)$ --- indicating it requires a claim on $\ell$ to execute safely, and gives up that claim.  Sequencing the acquisition and release would have effect $(\emptyset,\{\ell\})\rhd(\{\ell\},\emptyset)=(\emptyset,\emptyset)$.  Sequencing acquisitions for two locks $\ell_1$ and $\ell_2$ would have effect $(\emptyset,\{\ell_1\})\rhd(\emptyset,\{\ell_2\})=(\emptyset,\{\ell_1,\ell_2\})$, propagating the extra claim on $\ell_1$ that is not used by the acquisition of $\ell_2$.  This is true even when $\ell_1=\ell_2=\ell$ --- the overall effect would represent the recursive acquisition as two outstanding claims to hold $\ell$: $(\emptyset,\{\ell,\ell\})$.

A slightly more subtle example is the acquisition of a lock $\ell_2$ just prior to releasing a lock $\ell_1$, as would occur in the inner loop of hand-over-hand locking on a linked list: $(\textsf{acquire}\;\ell_2;\textsf{release}\;\ell_1)$ has effect $(\emptyset,\{\ell_2\})\rhd(\{\ell_1\},\emptyset)=(\{\ell_1\},\{\ell_2\})$.  The definition of $\rhd$ propagates the precondition for the release through the actions of the acquire; it essentially computes the minimal lock multiset required to execute both actions safely, and computes the final result of both actions' behavior on that multiset.

While use of sets rather than multisets would be simpler and would form an effect quantale for a given set of locks (with some use of disjoint union), such a formulation lacks an important property needed for substitution to behave correctly.  We introduce that property in Section \ref{sec:params}, and discuss subtle consequences of this in Section \ref{sec:concrete}.

\subsection{An Effect Quantale for Atomicity}
\label{sec:atomicity_quantale}
One of the best-known sequential effect systems is Flanagan and Qadeer's extension of \textsc{RCC/Java} to reason about atomicity~\cite{flanagan2003atomicity}, based on Lipton's theory of \emph{reduction}~\cite{lipton75} (called \emph{movers} in the paper).
The details of the movers are beyond what space permits us to explain in detail, but the essential ideas were developed for a simpler language and effect system in an earlier paper~\cite{flanagan2003tldi}, for which we give an effect quantale.

\begin{figure}[t]
\vspace{-0.5cm}
\begin{center}
$
\begin{minipage}{0.14\textwidth}
\begin{tikzpicture}
\node(Top){$\top$};
\node(A)[below of=Top]{$A$};
\node(R)[below right of=A]{$R$};
\node(L)[below left of=A]{$L$};
\node(B)[below right of=L]{$B$};
\draw(B)--(L);
\draw(B)--(R);
\draw(R)--(A);
\draw(L)--(A);
\draw(A)--(Top);
\end{tikzpicture}
\end{minipage}
\qquad
\begin{array}{|c|ccccc|}
\hline
; & B & L & R & A & \top\\
\hline
B & B & L & R & A & \top\\
R & R & A & R & A & \top\\
L & L & L & \top & \top & \top\\
A & A & A & \top & \top & \top\\
\top & \top & \top & \top & \top & \top\\
\hline
\end{array}
$
\end{center}
\vspace{-0.5cm}
\caption{Atomicity effects~\cite{flanagan2003tldi}: lattice and sequential composition.}
\label{fig:atomicity_lattice}
\vspace{-0.5cm}
\end{figure}
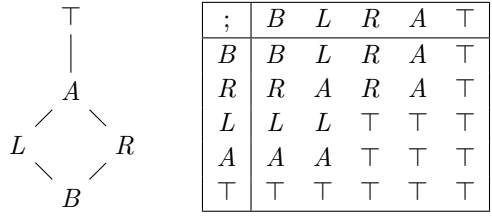

The core idea is that in a well-synchronized (i.e., data race free) execution, each action of one thread can be categorized by how it commutes with actions of other threads: a left ($L$) mover commutes left (earlier) with other threads' actions (e.g., a lock release), a right $R$ mover commutes later (e.g., lock acquire), a both $B$ mover commutes either direction (e.g., a well-synchronized field access).  A sequence of right movers, then both-movers, then left-movers \emph{reduces} to an atomic action ($A$).  Repeating the process wrapping movers around an atomic action can again reduce to an atomic action, verifying atomicity for even non-trivial code blocks including multiple lock acquisitions.  As a regular expression, any sequence of movers matching the regular expression $(R^*B^*)^*A(B^*L^*)^*$ reduces to an atomic action.  Effect trace fragments of this form demarcate expressions that evaluate as if they were physically atomic.

\begin{definition}[Atomicity Effect Quantale $\mathcal{A}$]
The effect quantale for Flanagan and Qadeer's simpler system~\cite{flanagan2003tldi} can be given as:
\begin{itemize}
\item $E=\{B,L,R,A,\top_{FQ},\mathsf{Err}\}$.  Note that $\top_{FQ}$ is the top effect in Flanagan and Qadeer's work --- their use does not itself require an error element.
\item $a\sqcup b$ is defined according to the lattice given by Flanagan and Qadeer~\cite{flanagan2003atomicity} (Figure \ref{fig:atomicity_lattice}) augmented with the new \textsf{Err} element as top (not shown in Figure \ref{fig:atomicity_lattice}).
\item $a\rhd b$ is defined according to Flangan and Qadeer's $;$ operator (Figure \ref{fig:atomicity_lattice}) plus our added \textsf{Err} element as an annihilator ($\mathsf{Err}\rhd a = \mathsf{Err} = a\rhd\mathsf{Err}$).
\item $\top=\mathsf{Err}$
\item $I=B$
\end{itemize}
\end{definition}

Flanagan and Qadeer also define an iterator operator on atomicities, used for ascribing effects to loops whose bodies have a particular atomicity.
We defer iteration until Section \ref{sec:iteration}, but will revisit this operator there.

Of course the atomicity effect quantale alone is insufficient to ensure atomicity, because atomicity
depends on correct synchronization.  The choice of effect for each program expression is not
insignificant, but full atomicity checking requires the \emph{product} of the synchronization and
atomicity effect quantales.  Thus, in Section \ref{sec:modeling2} we study an \emph{sequential
extension} to Flanagan and Qadeer's work using $\mathcal{L}\otimes\mathcal{A}$.

\begin{definition}[Products of Effect Quantales ($\otimes$)]
The product $Q\otimes R$ of effect quantales $Q$ and $R$ is given by the product of the respective
carrier lattices, with all pairs containing $\top$ from either constituent lattice merged into one
single \textsf{Err} element for the new lattice.  Other operations are lifted pointwise to each half
of the product, modified so if the lifting of an original operation from $Q$ or $R$ produces $\top$ in the respective lattice, the operation in the product produces \textsf{Err} (in the product lattice).  Identity is $(I_Q,I_R)$, and top is $\mathsf{Err}$.
\end{definition}

\subsection{Other Examples}
\label{sec:other_examples}
Our running example of tracking recursive lock ownership and atomicity is one of the better-known
sequential effect systems, but many more exist.
We are unaware of a source-level sequential effect system that \emph{does not} form an effect
quantale.

A particularly important class of these examples are those that reason specifically about execution traces.
All sequential effect systems reason to some degree about execution history, but some
examples from the literature have very expressive notions of the past.
Skalka's trace effects~\cite{Skalka2008} are (abstractions of) sets of event
traces, with trace concatenation (lifted to sets) as the monoid, and set union as join --- these
operations distributed as required by effect quantales, so adding a synthetic (unused) error element
to Skalka's work produces one.
Setting aside parallel composition (which we do not study), Nielson and Nielson's earlier communication
effect system for Concurrent ML~\cite{nielson1993cml} is similar to Skalka's.  Their \emph{behaviors}
act as trace set abstractions, with sequencing and non-deterministic choice (union) acting as an
effect quantale's monoid and join operations. (They also include a separate parallel composition of behaviors we do not model, discussed in Section \ref{sec:joinoids}.)  Their subtyping rules for behaviors imply the
required distributivity laws (though as with Skalka's system, we must add a synthetic error
element).
Similarly, Koskinen and Terauchi~\cite{Koskinen14LTR} use pairs of trace sets characterizing the
behavior of finite and infinite executions separately.  Their effects form an effect quantale, though
they additionally exploit set intersection for intersection effects.

\section{Iteration}
\label{sec:iteration}

Many sequential effect systems include a notion of iteration, used for constructs like explicit
loops.  The operator for this, usually written as a postfix ${}^*$, gives the overall effect of any
(finite\ifTR\footnote{Denotational approaches often require an additional notion of infinite
iteration --- ${}^\omega$ --- but for our syntactic proofs we only need ${}^*$ to model the finite
prefixes of an execution.}\fi) number of repetitions of an effect.

The iteration construct must follow from some fixed point construction on the semilattice.
However, the most obvious approach --- using a least fixed point theorem on effect quantales with a bottom element --- lacks an important property.
Instead, we detail an approach based on \emph{closure operators} on partially ordered sets in Section \ref{sec:closure_operator}, which applies to any effect quantale and coincides with manual iteration definitions for prior work.  First, in Section \ref{sec:iter_props}, we motivate a number of required properties for any derived notion of iterating an effect.

\subsection{Properties Required of an Iteration Operator}
\label{sec:iter_props}
Iteration operators must satisfy a few simple but important properties to be useful.  We first list, then explain these properties.\\
\centerline{
$
\begin{array}{l@{\quad}l@{\quad}l}
P1:  \forall e\ldotp e\sqsubseteq e^* &
P2:  \forall e\ldotp e\rhd e^* \sqsubseteq e^*~\textrm{and}~e^*\rhd e\sqsubseteq e^* &
P3:  \forall e\ldotp (e^*)^*=e^* \\
P4:  \forall e,f\ldotp (e\sqcup f)^* = e^*\sqcup f^* &
P5:  \forall e\ldotp I\sqsubseteq e^*
\end{array}
$
}

Property P1 ensures one iteration of a loop body has no greater effect than multiple iterations.  Similarly, the exact number of iterations should be immaterial (P2).  Nesting should not matter, since semantically the nested loop structure is dynamically unrolled to some number of sequential iterations (P3).  And P4 ensures certain equivalent ways of writing programs (e.g., a loop in each branch of a conditional vs. a conditional inside a loop) are effect-equivalent.  P1 and P2 are essential validity requirements for iteration. P3 and P4 are not strictly necessary, but permit many additional effect simplifications and figure prominently in prior work that found them important for building manageable effect systems~\cite{flanagan2003tldi,flanagan2003atomicity}.
P5 is slightly less obvious, but also critical: the least upper bound of the empty effect and some iterated effect should be the iterated effect.  This allows some helpful simplifications on effects (e.g., for a conditional whose only non-trivial branch contains a loop), but will play an essential role in the soundness proof later (the effect of not executing a loop is $I$).
This is also the property that fails for any straightforward use of least fixed point constructions --- all such constructions work on ascending chains rooted at $\bot$ (therefore requiring a bottom element), but unless $I$ is constrained to be $\bot$, there is no simple way to ensure with the fixed point equation alone that the resulting fixed point will be ordered above $I$.  Such a constraint is not unheard of ($\mathcal{A}$ satisfies it), but not universal.
$\mathcal{L}$ has no natural $\bot$, and adding a synthetic $\bot$ with identity behavior would mean $(\emptyset,\emptyset)$ would no longer be identity (it could be identity for all elements but $\bot$), making the role of the properties of the empty effect more difficult to characterize.

\subsection{Iteration via Closure Operators}
\label{sec:closure_operator}
For a general notion of iteration, we will use a \emph{closure operator} on a poset:
\begin{definition}[Closure Operator~\cite{birkhoff,blyth2006lattices,saraswat1991}]
A closure operator on a poset $(P,\sqsubseteq)$ is a function $f : P \rightarrow P$ that is
\begin{itemize}
    \item \textsf{Extensive}: $\forall e, e \sqsubseteq f(e)$
    \item \textsf{Idempotent}: $\forall e, f(f(e)) = f(e)$
    \item \textsf{Monotone}: $\forall e,e'\ldotp e\sqsubseteq e'\Rightarrow f(e)\sqsubseteq f(e')$
\end{itemize}
\end{definition}
Closure operators have several particularly useful properties~\cite{birkhoff,blyth2006lattices,saraswat1991}:
\begin{itemize}
\item Idempotence implies that the range of a closure operator is also the set of fixed points of the operator.
\item Closure operators on a poset are equivalent to their ranges.  In particular, from the range of a poset, we can recover the original closure operator by mapping each element of the poset to the least element of the range that is above that input.
\item A given subset of a poset is the range of a closure operator --- called a \emph{closure subset} --- if and only if for every element $x$ in the poset, every intersection with the principle up-set of $x$ ($x\uparrow=\{ y \mid x \sqsubseteq y \}$) has a bottom element~\cite[Theorem 1.8]{blyth2006lattices}.  (The left direction of the iff is in fact proven by constructing the closure operator as described above.)
\end{itemize}
This means that if we can identify the desired range of our iteration operation (the results of the iteration operator) and show that it meets the criteria to be a closure subset, the construction above will yield an appropriate closure operator, which we can take directly as our iteration operation.  For this to work, we must identify the desired range, and show it meets the requirements to induce a closure operator.

The natural choice is the set of elements for which sequential composition is idempotent, which we refer to as the \emph{freely iterable elements}:
\begin{definition}[Freely Iterable Elements]
The set of freely iterable elements $\mathsf{Iter}(Q)$ of an effect quantale $Q$ is defined as
$\mathsf{Iter}(Q) = \{ a\in Q \mid a\rhd a=a\}$.
\end{definition}
To induce a closure operator for this set, we must show it exists, and that it is in fact a closure subset.  The first is straightforward since \textsf{Err} and $I$ are freely iterable:
\begin{proposition}[Non-emptiness of Freely Iterable Elements]
\label{prop:nonempty_freely}
For any effect quantale $Q$, $\mathsf{Iter}(Q)$ is non-empty.
\end{proposition}
\ifTR
\begin{myproof}
Because effect quantales include an annihilating top element $\mathsf{Err}$ for sequencing, $\mathsf{Err}\rhd\mathsf{Err}=\mathsf{Err}$, so $\mathsf{Err}\in\mathsf{Iter}(Q)$.  Similarly, the identity element $I$ is among the freely iterable elements.
\end{myproof}
\fi

In general, the freely iterable elements do not themselves form a closure subset.
They could fail to form a closure subset in the case where some element $x$ is less than two incomparable freely iterable elements $y$ and $z$, but $x$ is not itself freely iterable and there is no other freely iterable element between --- there is no freely iterable $q$ such that $x\sqsubseteq q$ and $q\sqsubseteq y$ and $q\sqsubseteq z$.
Phrased differently the intersection of some element's principle up-set and the freely iterable elements lacks a least element.
However, this case appears uncommon; we have not observed it for any effect quantales we constructed, and cannot identify any systems in the literature with such irregular lattices.  So permitting closure operations only for effect quantales without this problem seems unproblematic in practice.

To derive our final solution, two further restrictions are required. First, the elements of our chosen closure subset must all reside at or above the identity in the semilattice, to ensure iteration permits loops to not execute.
Second, $\mathsf{Iter}(Q)$ must be closed under joins: $\forall x,y\in\mathsf{Iter}(Q)\ldotp (x\sqcup y)\in\mathsf{Iter}(Q)$.  This ensures iteration distributes over joins.
We call such effect quantales --- which have well-behaved closure operators --- \emph{iterable effect quantales}.

\begin{definition}[Iterable Effect Quantale]
An effect quantale $Q$ is \emph{iterable} if and only if for all $x$ the set
$x\uparrow \cap (I\uparrow\cap\mathsf{Iter}(Q))$
contains a least element and $\mathsf{Iter}(Q)$ is closed under joins.
\end{definition}
Another way to read the first part of the definition is that the closure operator will only exist for effect quantales for where, for every element $x$, if $x$ is $\sqsubseteq$ two incomparable freely iterable elements $y$ and $z$ (each greater than $I$), then there is some freely iterable element $q\sqsupset I$ such that $x\sqsubseteq q \sqsubseteq y$ and $q\sqsubseteq z$ (possibly $x$ itself).

\begin{proposition}[Closure for Iterable Effect Quantales]
\label{prop:closure_iterable}
For any iterable effect quantale $Q$, $I\uparrow\cap\mathsf{Iter}(Q)$ is a closure subset.
\end{proposition}
\begin{myproof}
$I\uparrow\cap\mathsf{Iter}(Q)$ is always non-empty, because $\mathsf{Iter}(Q)$ is non-empty and contains $\mathsf{Err}$ (Proposition \ref{prop:nonempty_freely}), and $I\sqsubseteq\mathsf{Err}$.
So if for every $x$, $x\uparrow\cap(I\uparrow\cap\mathsf{Iter}(Q))$ has a least element, $I\uparrow\cap\mathsf{Iter}(Q)$ is a closure subset~\cite[Theorem 1.8]{blyth2006lattices}.  This requirement is exactly the meaning of $Q$ being iterable, so this is a closure subset.
\end{myproof}
\begin{proposition}[Free Closure Operator on Iterable Effect Quantales]
For every iterable effect quantale $Q$, the function
$ F(X) \mapsto \mathsf{min}(X\uparrow\cap(I\uparrow\cap\mathsf{Iter}(Q))) $
is a closure operator satisfying properties P1--P5.
\end{proposition}
\ifTR
\begin{myproof}
That this is well-defined (i.e., the \textsf{min} in the definition exists) and that it is a closure operator follows immediately from Proposition \ref{prop:closure_iterable} and Blyth's construction~\cite[Theorem 1.8]{blyth2006lattices} of a closure operator from its range.
P1 follows from extensiveness of closure operators.
P2 follows from P1, isotonicity, and the fact that every element of $F$'s range is freely iterable.
P3 holds because the image of any closure operator is the set of its fixed points (alternatively because every freely iterable element --- like the result of the first iteration --- is self-iterating, and therefore its own closure for the second application of ${}^*$).
Setting aside P4, for a moment,
P5 holds by construction, as the closure subset for $F$ is constructed using only elements of $I$'s principal up-set ($I\uparrow$).

P4 holds as a consequence of simple facts about joins and principle up-sets, along with $\mathsf{Iter}(Q)$'s closure under joins.
Unfolding $F$, applying basic facts about principal up-sets, relying on the fact that the join of two elements of $\mathsf{Iter}(Q)$ is also in $\mathsf{Iter}(Q)$, and folding $F$:
\[\begin{array}{rcl}
a^*\sqcup b^* &=& \mathsf{min}(a\uparrow\cap(I\uparrow\cap\mathsf{Iter}(Q))) \sqcup \mathsf{min}(b\uparrow\cap(I\uparrow\cap\mathsf{Iter}(Q)))\\
&=& \mathsf{min}((a\sqcup b)\uparrow\cap(I\uparrow\cap\mathsf{Iter}(Q)))\\
&=& (a\sqcup b)^*
\end{array}
\]
\end{myproof}
\fi

\subsection{Iterating Concrete Effects}
We briefly compare the results of applying our derived iteration operation to effect quantales we have discussed to known iteration operations.
\begin{example}[Iteration for Atomicity]
The atomicity quantale $\mathcal{A}$ is iterable, so the free closure operator models iteration in that quantale.  The result is an operator that is the identity everywhere except for the atomic effect $A$, which is lifted to $\top_{FQ}$ when repeated (it is not an error, but no longer atomic).
This is precisely the manual definition Flanagan and Qadeer gave for iteration.  In Section \ref{sec:atomicity_quantale}, we claimed any trace fragment matching a regular expression evaluated as if it were physically atomic --- a property proven by Flanagan and Qadeer.  In terms of effect quantales, this is roughly equivalent to the claim that $(R^*\rhd B^*)^*\rhd A\rhd(B^*\rhd L^*)^*\sqsubseteq A$.  With our induced iteration operator, this has a straightforward proof:
\[
\begin{array}{rll}
(R^*\rhd B^*)^*\rhd A\rhd(B^*\rhd L^*)^* & =(R\rhd B)^*\rhd A\rhd(B\rhd L)^* & \textrm{since $R^*=R$, $B^*=B$}\\
                                         & =R^*\rhd A\rhd L^* & \textrm{$B$ is unit for $\rhd$}\\
                                         & =R\rhd A\rhd L  & \textrm{since $R^*=R$, $B^*=B$}\\
                                         & = A & \textrm{by definition of $\rhd$}
\end{array}
\]
\end{example}

\begin{example}[Iteration for Commutative Effect Quantales]
For any bounded join semilattice, we have by Lemma \ref{lem:subsume} a corresponding effect quantale that reuses join for sequencing (and thus, $\bot$ for unit), making the sequencing operation commutative.  For purposes of iteration, this immediately makes all instances of this free effect quantale iterable, as idempotency of join ($x\sqcup x=x$) makes all effects freely iterable.  The resulting iteration operator is the identity function, which exactly models the standard type rule for imperative loops in commutative effect systems, which reuse the effect of the body as the effect of the loop:
\[
\inferrule{\Gamma\vdash e_1 : \mathsf{bool} : \chi_1\\\Gamma\vdash e_2 : \mathsf{unit} \mid \chi_2}{\Gamma\vdash \mathsf{while}(e_1)\{ e_2 \} : \mathsf{unit} \mid \chi_1\sqcup\chi_2 }
\]
For a quantales where sequencing is merely the join operation on the semilattice, the above standard rule can be derived from our rule in Section \ref{sec:soundness} by simplifying the result effect:
\[
\chi_1\rhd(\chi_2\rhd\chi_2)^*=\chi_1\rhd(\chi_2\rhd\chi_2)=\chi_1\sqcup(\chi_2\sqcup\chi_2)=\chi_1\sqcup\chi_2
\]
\end{example}

\begin{example}[Loop Invariant Locksets]
For the lockset effect quantale $\mathcal{L}$, the freely iterable elements are all actions that do not acquire or release any locks --- those of the form $(a,a)$ for some $a$, and $\top$.  These are isomorphic to the set of all multisets formed over the set of locks (plus the error element $\top$), and for those elements the join is equivalent to the complete lattice under multiset inclusion (again, plus the top error element).  Since $I$ is $(\emptyset,\emptyset)$ (which has no elements below it),
$ I\uparrow\cap\mathsf{Iter}(\mathcal{L})=I\uparrow\cap(\{ (a,b) \mid a=b \}\cup\{\top\})=\{ (a,b) \mid a=b \}\cup\{\top\} $.
Because the freely iterable elements above unit form a \emph{complete} lattice, $\mathcal{L}$ is iterable.
The resulting closure operator is the identity on the freely iterable elements, and takes all actions that acquire or release locks to $\top$ (\textsf{Err}).  This is exactly what intuition suggests as correct --- the iterable elements are those that hold the same locks before and after each loop iteration, and attempts to repeat other actions should be errors.
\end{example}

\section{Syntactic Type Soundness for Generic Sequential Effects}
\label{sec:soundness}
In this section we give a purely syntactic proof that effect quantales are adequate for syntactic soundness proofs of sequential type-and-effect systems.  For the growing family of algebraic characterizations of sequential effect systems, this is the first soundness proof we know of that is (1) purely syntactic, (2) handles the indexed versions of the algebra required for singleton effects, (3) addresses effect polymorphism, and (4) includes direct iteration constructs.  This development both more closely mirrors common type soundness developments for applied effect systems than the category theoretic approaches discussed in Section \ref{sec:semantics}, and demonstrates machinery which would need to be developed in an analagous way for syntactic proofs using those concepts.  Thus, for hypothetical future effect systems requiring more structure than effect quantales provide, our techniques provide guidance on using those more general concepts without switching to category theoretic denotational semantics.

We give this soundness proof for an \emph{abstract} effect system --- primitive operations, the notion of state, and the overall effect systems are all abstracted by a set of parameters (operational semantics for primitives that are aware of the state choice).
This alone requires relatively little mechanism at the type level, but we wish to not only demonstrate that effect quantales are sound, but also that they are adequate for non-trivial existing sequential effect systems.  In order to support such embeddings (see Sections \ref{sec:modeling} and \ref{sec:modeling2}), the type system includes parametric polymorphism --- over types and effects as different kinds --- as well as singleton types (e.g., for reference types with region tags) and effect constructors (such as effects mentioning particular locks).  We consider effects equal according to the equations induced by effect quantale properties, and for families of effects indexed by values we identify the families with uses of effect constructors applied to singleton types.
We demonstrate embeddings by directly translating equivalent constructs, and building artificial terms to model other constructs.  These artificial terms' \emph{derived} type rules directly match the language we embed, though the dynamic semantics may not be preserved (for example, we do not model concurrency).  While unsuitable for a general framework in the style of a language workbench, this is adequate to show that our characterization of sequential effect systems' structure is flexible.

The language we study includes no built-in means to introduce a non-trivial (non-identity) effect, relying instead on the supplied primitives.
The language also includes only the simplest form of parametric polymorphism for effects (and types), without bounding, constraints~\cite{Grossman2002Cyclone}, relative effect declarations~\cite{vanDooren2005,rytz12}, qualifier-based effects~\cite{ecoop13}, or any other richer forms of polymorphism.  Our focus is demonstrating compatibility of effect quantales with effect polymorphism and singleton effects, rather than to build a particularly powerful framework.

We stage the presentation to first focus on core constructs related to effect quantales, then briefly recap machinery from Systems F and F$\omega$ (and small modifications beyond what is standard), before proving type soundness.  Section \ref{sec:modeling2} gives an embedding from Flanagan and Qadeer's sequential effect system for atomicity~\cite{flanagan2003tldi} into our core language to establish that it is not only sound, but expressive.

\subsection{Parameters to the Langauge}
\label{sec:params}
We parameterize our core language by a number of external features.
First among these, is a slight extension of an effect quantale --- an \emph{indexed} effect quantale.
\begin{definition}[Indexed Effect Quantale]
An indexed effect quantale is a quantale whose elements (and therefore operations) are parameterized over some set.\footnote{For those accustomed to typed meta-logics (e.g., \textsc{Coq}), one could view an indexed quantale as roughly the type $\forall\alpha:\mathsf{Type}\ldotp\{\mathsf{Decidable}\; \alpha\}\rightarrow\alpha\rightarrow\mathsf{Quantale}$.  The point is that the details of the set are irrelevant to the quantale's definition.}
\end{definition}
The lock set effect quantale $\mathcal{L}$ we described earlier is in fact an indexed effect quantale, parameterized by the set of lock names to consider.

Because the set of well-typed values changes during program execution, we will need to transport terms well-typed under one use of the quantale into another use of the quantale, under certain conditions.  The first is the introduction of new well-typed values (e.g., from allocating a new heap cell), requiring a form of inclusion between indexed quantales.  The second is due to substitution: our language considers variables to be values, but during substitution some variable may be replaced by another value that was already present in the set.  This essentially collapses what statically appears as two values into a single value, thus \emph{shrinking} the set of values distinguished inside the quantale.  Each requires a different kind of homomorphism between effect quantales, with different properties.
\begin{definition}[Effect Quantale Homomorphism]
An \emph{effect quantale homomorphism} between two effect quantales $Q$ and $R$ is a join semilattice homomorphism (a function between the carrier sets that preserves joins) that additionally preserves sequencing.
\end{definition}
\begin{definition}[Monotone Indexed Effect Quantale]
An indexed effect quantale $Q$ is called \emph{monotone} when for two sets $S$ and $T$ where $S\subseteq T$, the inclusion function from the carrier of $Q(S)$ to the carrier set of $Q(T)$ induces the obvious inclusion homomorphism.
\end{definition}
\begin{definition}[Collapsible Indexed Effect Quantale]
An indexed effect quantale $Q$ is called \emph{collapsible} when for any non-empty set $S$ and additional element $x$ (not in $S$), a function $f$ from $S\cup\{x\}$ to $S$ that is the identity on elements of $S$ induces a corresponding homomorphism where only sequences and joins that produced $\top$ under $S\cup\{x\}$ produce $\top$ when transported by the homomorphism (i.e.,
$f(a)\rhd f(b)=\top\Rightarrow a\rhd b=\top$, similarly for joins).
\end{definition}
We parameterize our core language by a monotone, collapsible indexed effect quantale $Q$.  Monotonicity is a natural requirement, but collapsibility has some subtle consequences we defer to Section \ref{sec:relwork}.
Any constant (i.e., non-indexed) effect quantale trivially lifts to a monotone collapsible indexed effect quantale that ignores its arguments.  The product construction $\otimes$ lifts in the expected way.

The language parameters also include:

\begin{itemize}
\item An abstract notion of state, usually noted by $\sigma\in\mathsf{State}$.  For a pure calculus \textsf{State} might be unit, while other languages might instantiate it to a heap, etc.
\item A set of primitives $p_i$ operating on terms and \textsf{State}s.  This includes modeling additional values that do not interact directly with general terms, such as references.
\item A set of type families $T_i$ for describing the types of primitives.
\item A meta-function $K$ for ascribing appropriate kinds to types in $T_i$.  Thus, reference types may be modeled this way.
\item A meta-function $\delta$ for ascribing a type to some primitive that is independent of the state --- i.e., source-level primitive operations (but not store references).  $\delta$ is constrained such that for values whose types are applicative (i.e., function types and quantified types) only the very last such type may have non-unit effect.
\item A partially ordered state type environment $\Sigma\in\mathsf{StateEnv}$, which maps a subset of the primitives to types.  The least element in the partial order is $\delta$ (used for source typing of primitives).
\item A \emph{partial} primitive semantics $\llbracket-\rrbracket : \mathsf{Term}\rightharpoonup\mathsf{State}\rightarrow \mathsf{Term}\times E\times\mathsf{State}$ defined only on full applications of primitive operations (i.e., fully-applied primitive operations, judged according to the types from $\delta$).
\end{itemize}
For type soundness, we will rely on the following:
\begin{itemize}
\item Types produced by $\delta$ must be well-formed in the empty environment, and must not be closed base types (e.g., the primitives cannot add a third boolean, which would break the canonical forms lemma).
\item Effects produced by $\llbracket-\rrbracket$ are valid for the quantale paramerized by the values at the call site (i.e., the dynamic effects depend only on the values at the call).
\item There is a relation $Q\vdash \sigma : \Sigma$ for well-typed states.
\item When the primitive semantics are applied to well-typed primitive applications and a well-typed state, the resulting term is well-typed (in the empty environment) with argument substitutions applied, and the resulting state is well-typed under some ``larger'' state type:\\
$\begin{array}{l}
\epsilon;\Sigma\vdash p_i\;\overline{v} : \tau \mid \gamma \land Q\vdash \sigma:\Sigma \land \llbracket p_i\;\overline{v}\rrbracket(\sigma)=(v',\gamma',\sigma') \\
\qquad\Rightarrow \exists \Sigma'\ldotp \Sigma\le\Sigma' \land \epsilon;\Sigma'\vdash v' : \tau[\overline{v}/\mathsf{args}(\delta(p_i))] \mid I \land Q\vdash\sigma':\Sigma'
\end{array}$\\
We call this property \emph{primitive preservation}.
\end{itemize}

This setup leads to a delicate dependency order among these parameters and the core language to avoid circularity.  Such circularity is manageable with sophisticated tools in the ambient logic~\cite{Delaware2013MLC,birkedal2013intensional}, but we prefer to avoid them for now.
The parameters and language components are stratified as follows:
\begin{itemize}
    \item The syntax of kinds is closed.
    \item The core language's syntax for terms and types is mutually defined (the language contains explicit type application and singleton types), parameterized by $T_i$ and $p_i$.  The latter parameters are closed sets, to the mutual definition is confined to the core.
    \item The type judgment depends on (beyond terms, types, and kinds) $\delta$, $K$, and $\mathsf{StateEnv}$.
    \item $\mathsf{State}$ may depend on terms, types, and kinds.
    \item The dynamic semantics will depend on terms, types, kinds, \textsf{State}, and $\llbracket-\rrbracket$ (which cannot refer back to the main dynamic semantics).
    \item Primitive preservation depends on the typing relation and state typing.
    \item The type soundness proof will rely on all core typing relations, state typing (which may be defined in terms of source typing), and the primitive preservation property.
\end{itemize}
Ultimately this leads to a well-founded set of dependencies for the soundness proof.

\subsection{The Core Language, Formally}

\begin{figure}[t!]
\small
\[\begin{array}{lrcl}
\textsf{Kinds} & \kappa & ::= & \star \mid \mathcal{E} \mid \kappa\Rightarrow\kappa\\
\textsf{Types} & \tau & ::= & T_i \mid \tau\;\tau \mid E_Q \mid \Pi x:\tau\overset{\tau}{\rightarrow}\tau \mid \alpha \mid \mathsf{bool} \mid \forall\alpha::\kappa\overset{\tau}{\rightarrow}\tau \mid \mathsf{unit} \mid \mathcal{S}(v) \\
\mathsf{Terms} & e & ::= & p_i \mid (\lambda x\ldotp e) \mid e\;e\mid x \mid \mathsf{true} \mid \mathsf{false} \mid \mathsf{if}\;e\;e\;e \mid \mathsf{while}\;e\;e \mid (\Lambda \alpha::\kappa\ldotp e) \mid e[\tau] \mid ()\\
\mathsf{TypeEnv} & \Gamma & ::= & \epsilon \mid \Gamma,x:\tau \mid \Gamma,\alpha::\kappa\\
\mathsf{Values} & v & ::= & p_i \mid (\lambda x\ldotp e) \mid x \mid \mathsf{true} \mid \mathsf{false}
\end{array}\]
\begin{mathpar}
\fbox{$\vdash\Gamma$}
\and
\inferrule{ }{\vdash\epsilon}
\and
\inferrule{\Gamma\vdash\tau::\star\\x\not\in\Gamma}{\vdash\Gamma,x:\tau}
\and
\inferrule{\alpha\not\in\Gamma}{\Gamma\vdash\alpha::\kappa}
\\
\fbox{$\Gamma\vdash \tau :: \kappa$}
\and
\inferrule{ }{\Gamma\vdash T_i :: K(T_i)}
\and
\inferrule{\Gamma(\alpha)=\kappa}{\Gamma\vdash\alpha::\kappa}
\and
\inferrule{
    \Gamma\vdash\tau::\kappa\Rightarrow\kappa'\\
    \Gamma\vdash\tau'::\kappa
}{
    \Gamma\vdash\tau\;\tau'::\kappa'
}
\and
\inferrule{E\in Q(\Gamma) }{\Gamma\vdash E :: \mathcal{E}}
\and
\inferrule{
    \Gamma\vdash\tau::\star\\\\
    \Gamma,x:\tau\vdash\gamma::\mathcal{E}\\\\
    \Gamma,x:\tau\vdash\tau'::\star
}{
    \Gamma\vdash(\Pi x:\tau\overset{\gamma}{\rightarrow}\tau') :: \star
}
\and
\inferrule{ }{\Gamma\vdash\mathsf{bool}::\star}
\and
\inferrule{ }{\Gamma\vdash\mathsf{unit}::\star}
\and
\inferrule{\Gamma\vdash v : \tau \mid I}{\Gamma\vdash \mathcal{S}(v) :: \star}
\and
\inferrule{
    \Gamma,\alpha::\kappa\vdash\gamma::\mathcal{E}\\\\
    \Gamma,\alpha::\kappa\vdash\tau :: \star
}{
    \Gamma\vdash \forall\alpha::\kappa\overset{\gamma}{\rightarrow}\tau :: \star
}
\\
\fbox{$\Gamma\vdash e : \tau \mid \gamma$}
\and
\inferrule{ }{\Gamma\vdash p_i : \delta(p_i) \mid I}
\and
\inferrule{\Gamma(x)=\tau}{\Gamma\vdash x : \tau \mid I}
\and
\inferrule{
    \Gamma,x:\tau\vdash e : \tau' \mid \gamma_e \\ \gamma_e\sqsubseteq\gamma
}{
    \Gamma\vdash(\lambda x\ldotp e) : \Pi x:\tau\overset{\gamma}{\rightarrow}\tau' \mid I
}
\and
\inferrule{
    \Gamma\vdash e_1 : \Pi x:\tau\overset{\gamma}{\rightarrow}\tau' \mid \gamma_1\\
    \Gamma\vdash e_2 : \tau \mid \gamma_2\\
    x\not\in\mathsf{FV}(\gamma,\tau')\vee\mathsf{Value}(e_2)
}{
    \Gamma\vdash e_1\;e_2 : \tau'[e_2/x] \mid \gamma_1\rhd\gamma_2\rhd\gamma[e_2/x]
}
\and
\inferrule{b\in\{\mathsf{true},\mathsf{false}\}}{\Gamma\vdash b : \mathsf{bool} \mid I}
\and
\inferrule{
    \Gamma\vdash c : \bool \mid \gamma_c\\
    \Gamma\vdash e_1 : \tau \mid \gamma_1\\
    \Gamma\vdash e_2 : \tau \mid \gamma_2
}{
    \Gamma\vdash\mathsf{if}\;c\;e_1\;e_2 : \tau \mid \gamma_c\rhd(\gamma_1\sqcup\gamma_2)
}
\and
\inferrule{
    \Gamma\vdash c : \mathsf{bool} \mid \gamma_c\\
    \Gamma\vdash e : \tau \mid \gamma_b
}{
    \Gamma\vdash\mathsf{while}\;c\;e : \mathsf{unit} \mid \gamma_c\rhd(\gamma_b\rhd\gamma_c)^*
}
\and
\inferrule{
    \Gamma,\alpha::\kappa\vdash e : \tau \mid \gamma
}{
    \Gamma\vdash(\Lambda\alpha::\kappa\ldotp e) : \forall\alpha::\kappa\overset{\gamma}{\rightarrow}\tau \mid I
}
\and
\inferrule{
    \Gamma\vdash e : \forall\alpha::\kappa\overset{\gamma}{\rightarrow}\tau \mid \gamma_e\\
    \Gamma\vdash \tau' :: \kappa
}{
    \Gamma\vdash e[\tau'] : \tau[\tau'/\alpha] \mid \gamma_e\rhd\gamma[\tau'/\alpha]
}
\and
\inferrule{ }{\Gamma\vdash () : \mathsf{unit} \mid I }
\\
\fbox{$\sigma,e\rightarrow^{\gamma}_Q \sigma',e'$}
\quad
\inferrule{ }{\sigma,(\lambda x\ldotp e)\;v\rightarrow^I_Q \sigma,e[v/x]}
\quad
\inferrule{ }{\sigma,(\Lambda\alpha::\kappa\ldotp e)[\tau]\rightarrow^I_Q e[\tau/\alpha]}
\quad
\inferrule{
    \llbracket p_i\;\overline{v}\rrbracket(\sigma)=(e',\gamma,\sigma')
}{
    \sigma,p_i\;\overline{v}\rightarrow^{\gamma}_Q \sigma',e'
}
\and
\inferrule{ }{\sigma,\mathsf{if}\;\mathsf{true}\;e_1\;e_2\rightarrow^I_Q \sigma,e_1}
\quad
\inferrule{ }{\sigma,\mathsf{if}\;\mathsf{false}\;e_1\;e_2\rightarrow^I_Q \sigma,e_2}
\quad
\inferrule{ }{\sigma,\mathsf{while}\;e\;e_b\rightarrow^I_Q \sigma,\mathsf{if}\;e\;(e_b;\mathsf{while}\;e\;e_b)\;()}
\\
\fbox{$\sigma,e\overset{\gamma}{\longrightarrow}^*_Q \sigma',e'$}
\and
\inferrule{ }{\sigma,e\overset{I}{\longrightarrow}^*_Q\sigma,e}
\and
\inferrule{
    \sigma,e\overset{\gamma}{\longrightarrow}^*_Q \sigma',e'\\
    \sigma',e'\rightarrow^{\gamma'}_Q \sigma'',e''
}{\sigma,e{\xrightarrow{\gamma\rhd\gamma'}}^*_Q\sigma'',e''}
\end{mathpar}
\caption{A generic core language for sequential effects, omitting straightforward structural rules from the operational semantics. $;$ is standard sugar for sequencing with in a CBV lambda calculus.}
\label{fig:syntax}
\end{figure}

Figure \ref{fig:syntax} gives the (parameterized) syntax of kinds, types, and terms.  Most of the structure should be familiar from standard effect systems and Systems F and F$\omega$ (with multiple kinds, as in the original polymorphic effect calculus~\cite{lucassen88}), plus standard while loops and conditionals with effects sequenced as in Section \ref{sec:bg}.  We focus on the differences.

The language includes a dependent product (function) type, which permits program values to be used in types and effects.  This is used primarily through effects --- elements of an effect quantale may mention elements of the set --- and through the singleton type constructor $\mathcal{S}(-)$, which associates a type (classifying no terms) with each program value.  Use of the dependent function space is restricted to syntactic values (which includes variables in our call-by-value language) --- the application rule requires that either the argument is a syntactic value, or the function type's named argument does not appear in the effect or result type.  In the latter case, for concrete types we will use the standard $\tau\overset{\gamma}{\rightarrow}\tau'$ notation.  A minor item of note is that dependent function types and quantified types bind their argument in the function's effect as well as in the result type.  This permits uses such as a function acquiring the lock passed as an argument.
One small matter important to the soundness proof: for any value, the effect of the value itself is the identity effect $I$.

Every rule carries an implicit side condition that the resulting effect is $\neq\top$.  Since $\top$ acts as the error element, this permits effect systems to completely reject certain event orders.

A slightly more subtle point concerns the kinding judgment for effects.  The requirement is that an effect $E$ is valid if it is contained in $Q(\Gamma)$.  This is because the type system is actually given with respect to an indexed effect quantale, as described above, which accepts some set to parameterize the system by.  $Q(\Gamma)$ is $Q$ instantiated with the set of well-typed values under $\Gamma$.

The operational semantics is mostly standard: a labelled transition system over pairs of states and terms, where the label is the effect of the basic step.  We omit the structural rules that simply reduce a subexpression and propagate state changes and the effect label in the obvious way.
The only other subtlety of the single-step relation is that when reducing invocations of primitives, if a primitive's semantics via $\llbracket-\rrbracket$ are defined only on larger-arity calls than what has been reduced to values $\overline{v}$ (which also includes type applications), the next argument applied is reduced, structurally.  Incomplete applications of primitives remain stuck.
We also give a transitive reduction relation $\overset{\gamma}{\longrightarrow}^*_Q$ which accumulates the effects of each individual step.

\paragraph*{Runtime Typing}
Figure \ref{fig:syntax} gives the source type system.  For the runtime type system, three changes are made.  First, a state type $\Sigma$ is added to the left side of each judgment in the standard way.  Second, primitive typing is changed to rely on $\Sigma$ rather than $\delta$ (recall that $\delta$ is the least element in the partial order, so all $\Sigma$ will extend $\delta$).  And third, the effect kinding is modified to check for effects in $Q(\Gamma,\Sigma)$ --- the effect quantale instantiated for a set of values well-typed under $\Gamma$ and $\Sigma$, allowing values introduced at runtime (such as dynamically allocated locks or references) to appear in effects.

\subsection{Syntactic Safety}
Syntactic type safety proceeds in the normal manner (for a language with mutually-defined types and terms), with only a few wrinkles due to effect quantales.
\ifTR
Here we give the major lemmas involved in the type safety proof, with outlines of the proofs themselves.
\fi

Substitution lemmas are proven by induction on the expression's type derivation, exploiting the fact that all values' effects before subeffecting are $I$:
\begin{lemma}[Term Substitution]
\label{lem:term_subst}
If $\Gamma,x:\tau\vdash e : \tau \mid \gamma$ and $\Gamma\vdash v : \tau \mid I$, then $\Gamma\vdash e[v/x] : \tau[v/x] \mid \gamma[v/x]$,
and simultaneously
if $\Gamma,x:\tau\vdash \tau' :: \kappa$ and $\Gamma\vdash v : \tau \mid I$ then $\Gamma\vdash\tau'[v/x] :: \kappa$.
\end{lemma}
\begin{myproof}
By simultaneous induction on the typing and kinding relations.
The only subtle case is substitution of a variable occurring in an effect.
In this case, the set of well-typed values is being reduced in size by one, with uses of the substituted variable being replaced by the new value.  This induces the type of homomorphism relevant for collapsible (indexed) effect quantales.  By assumption $Q$ is collapsible, so applying the appropriate homomorphism as substitution yields an effect that is well-kinded in the smaller type environment.
\end{myproof}

\ifTR
\begin{lemma}[Type Substitution]
\label{lem:type_subst}
If $\Gamma,\alpha::\kappa\vdash e : \tau \mid \gamma$ and $\Gamma\vdash \tau' :: \kappa$, then $\Gamma\vdash e[\tau'/\alpha] : \tau[\tau'/\alpha] \mid \gamma[\tau'/\alpha]$, and simultaneously
if $\Gamma,\alpha::\kappa\vdash \tau' :: \kappa'$ and $\Gamma\vdash \tau'' :: \kappa$, then $\Gamma\vdash\tau'[\tau''/\alpha]::\kappa'$.
\end{lemma}
\begin{myproof}
By simultaneous induction on the typing and kinding relations.  Because types (and effects) may not appear inside effects, replacing a type in an effect is a no-op, and no special treatment is required.
\end{myproof}
\begin{lemma}[Canonical Forms]
\label{lem:canonical}
If $\epsilon\vdash v : \tau \mid \gamma$ then:
\begin{itemize}
\item If $\tau=\Pi x:\tau'\overset{\gamma'}{\rightarrow}\tau''$, then $v$ is a primitive ($p_i$) or $v$ is of the form $(\lambda y\ldotp e)$ and $I\sqsubseteq\gamma$.
\item If $\tau=\forall\alpha::\kappa\overset{\gamma'}{\rightarrow}\tau'$ then $v$ is a primitive or $v$ is of the form $(\Lambda\alpha::\kappa\ldotp e)$.
\item If $\tau=\mathsf{bool}$, $v=\mathsf{true}\vee v=\mathsf{false}$.
\item If $\tau=\mathsf{unit}$, $v=()$.
\end{itemize}
\end{lemma}
\begin{myproof}
    Entirely standard, with the exception that the boolean and unit cases rely on $\delta$'s restriction to not give any primitives those closed base types.
\end{myproof}
\fi

We give type preservation below, assuming an iterable effect quantale.  This assumption is only used in \textsf{while}-related cases, so this proof also shows soundness for programs without loops under non-iterable quantales.
\begin{lemma}[One Step Type Preservation]
\label{lem:onestep_preservation}
For all $Q$, $\sigma$, $e$, $e'$, $\Sigma$, $\tau$, $\gamma$, and $\gamma'$, if
$\epsilon;\Sigma\vdash e : \tau \mid \gamma$,
$Q\vdash \sigma : \Sigma$,
$\delta\le\Sigma$, and
$\sigma,e\rightarrow^{\gamma'}_Q \sigma',e'$
then there exist $\Sigma'$, $\gamma''$ such that
$\epsilon;\Sigma'\vdash e' : \tau \mid \gamma''$,
$Q\vdash\sigma':\Sigma'$,
$\Sigma\le\Sigma'$,
$\gamma'\rhd\gamma''\sqsubseteq\gamma$.
\end{lemma}
\begin{myproof}
By induction on the reduction relation.
\ifTR
We present only the non-trivial reducts (i.e., we omit structural rules that simply reduce a subterm and propagate state and effect labels).
\fi
\begin{itemize}
\ifTR
\item Case \textsc{E-App}: Here we know $e=(\lambda x\ldotp e_b)\;v$, $e'=e_b[v/x]$, $\sigma=\sigma'$ and $\gamma'=I$.  By inversion on the typing derivation:
\[\begin{array}{l@{\qquad}l}
\epsilon;\Sigma\vdash(\lambda x\ldotp e_b):\Pi x:\tau_{arg}\overset{\gamma_f}{\rightarrow}\tau_{res} \mid \gamma_a
&
\tau=\tau_{res}[v/x]
\\
\epsilon;\Sigma\vdash v : \tau_{arg}\mid\gamma_v
&
x\not\in\mathsf{FV}(\gamma_f,\tau_{res})\vee\mathsf{Value}(v)
\\
\gamma_a\rhd\gamma_v\rhd(\gamma_f[v/x])=\gamma
\end{array}
\]
By inversion on value typing, $\gamma_a=\gamma_v=I$, so $(\gamma_f[v/x])=\gamma$.  Also from the inversion on the function's type derivation, $\epsilon,x:\tau_{arg};\Sigma\vdash e_b : \tau_{res} \mid \gamma_e$ and $\gamma_e\sqsubseteq\gamma_f$.
By term substitution (Lemma \ref{lem:term_subst}) we then have $\epsilon;\Sigma\vdash e_b[v/x] : \tau_{res}[v/x] \mid \gamma_e[v/x]$.
We then know $I\rhd\gamma_e[v/x]\sqsubseteq\gamma_f[v/x]\sqsubseteq\gamma$, and as the state did not change, $\Sigma'=\Sigma$ and the state remains well-typed.
\item Case \textsc{E-TApp}: Here $e=(\Lambda\alpha::\kappa\ldotp e_b)[\tau_\alpha]$, $\gamma'=I$, $e'=e_b[\tau_\alpha/\alpha]$, $\sigma=\sigma'$.
By inversion on the typing assumption:
\[\begin{array}{l@{\qquad}l}
\epsilon;\Sigma\vdash(\Lambda\alpha::\kappa\ldotp e_b) : \forall\alpha::\kappa\overset{\gamma_f}{\rightarrow}\tau_{res}\mid\gamma_e
&
\tau=\tau_{res}[\tau_\alpha/\alpha]
\\
\epsilon;\Sigma\vdash \tau_\alpha :: \kappa
&
\gamma=\gamma_e\rhd\gamma_f[\tau_\alpha/\alpha]
\end{array}\]
By inversion on value typing, $\gamma_e=I$, so $\gamma=\gamma_f[\tau_\alpha/\alpha]$.
Also from the inversion on the function's type derivation, $\epsilon,\alpha::\kappa;\Sigma\vdash e_b : \tau_{res} \mid \gamma_f$.
By type substitution (Lemma \ref{lem:type_subst}) we then have $\epsilon;\Sigma\vdash e_b[\tau_\alpha/\alpha] : \tau_{res}[\tau_\alpha/\alpha] \mid \gamma_f[\tau_\alpha/\alpha]$.  We also know $I\rhd\gamma_f[\tau_\alpha/\alpha]=\gamma_f[\tau_\alpha/\alpha]\sqsubseteq\gamma$, and as the state did not change, $\Sigma'=\Sigma$ and the state remains well-typed.
\item Case \textsc{E-Prim}: Here $e=p_i\;\overline{v}$, and $\llbracket p_i\;\overline{v}\rrbracket(\sigma)=(e',\gamma',\sigma')$.
By inversion on typing, $\overline{v}$ is a series of values and types (possibly including effects) passed as successive arguments, where later arguments' types may depend on earlier arguments' values.  Further, the length and component make-up are consistent with the type ascribed to $p_i$ by $\delta$, and by constraints on $\delta$\footnote{Recall that only the last effect of a curried function may be non-unit.} and value typing, the overall effect is the final effect of the type ascribed by $\delta$:
\[ \epsilon;\Sigma\vdash p_i\;\overline{v} : \mathsf{LastResult}(\delta(p_i))[\overline{v}/\mathsf{args}(\delta(p_i))] \mid \mathsf{LastEffect}(\delta(p_i))[\overline{v}/\mathsf{args}(\delta(p_i))] \]
By primitive preservation, the result of any such operation must produce a valid new state, $e'$ is some value $v'$, and the produced effect $\gamma'=\mathsf{LastEffect}(\delta(p_i))[\overline{v}/\mathsf{args}(\delta(p_i))]$.
Also by primitive preservation, there must exist a new $\Sigma'$ ordered after $\Sigma$, and $\epsilon;\Sigma'\vdash v' : \mathsf{LastResult}(\delta(p_i))[\overline{v}/\mathsf{args}(\delta(p_i))] \mid I$.

\item Case \textsc{E-IfTrue}: Here $e=\mathsf{if}\;\mathsf{true}\;e_1\;e_2$, $e'=e_1$, $\sigma=\sigma'$, $\gamma'=I$.  By inversion on typing:
\[\begin{array}{l@{\qquad}l}
\epsilon;\Sigma\vdash\mathsf{true}:\mathsf{bool}\mid\gamma_{\mathsf{true}}
&
\epsilon;\Sigma\vdash e_1 : \tau \mid \gamma_1
\\
\epsilon;\Sigma\vdash e_2 : \tau \mid \gamma_2
&
\gamma=\gamma_{\mathsf{true}}\rhd(\gamma_1\sqcup\gamma_2)
\end{array}\]
By inversion on value typing, $\gamma_\mathsf{true} = I$.
By local hypothesis, immediately $\epsilon;\Sigma\vdash e_1 : \tau \mid\gamma_1$, and by effect quantale laws $I\rhd\gamma_2\sqsubseteq(\gamma_1\sqcup\gamma_2)$.
State is unchanged, and remains well-typed under $\Sigma'=\Sigma$.
\item Case \textsc{E-IfFalse}: Analagous to \textsc{E-IfTrue}.
\fi
\item Case \textsc{E-While}: Here $e=\mathsf{while}\;e_c\;e_b$, $\gamma'=I$, $\sigma=\sigma'$, and $e'=\mathsf{if}\;e_c\;(e_b;(\mathsf{while}\;e_c\;e_b))\;()$.
By inversion on typing:
\[\begin{array}{l@{\qquad}l@{\qquad}l@{\qquad}l}
\epsilon;\Sigma\vdash e_c:\mathsf{bool} \mid \gamma_c
&
\epsilon;\Sigma\vdash e_b:\tau_b \mid \gamma_b
&
\gamma=\gamma_c\rhd(\gamma_b\rhd\gamma_c)^*
&
\tau=\mathsf{unit}
\end{array}\]
By \textsc{T-If}, \textsc{T-Unit}, desugaring $;$ to function application, and weakening, \\\mbox{$\epsilon;\Sigma\vdash\mathsf{if}\;e_c\;(e_b;(\mathsf{while}\;e_c\;e_b))\;() : \mathsf{unit} \mid \gamma_c\rhd(((\gamma_b\rhd\gamma_c)\rhd(\gamma_b\rhd\gamma_c)^*)\sqcup I)$}.
State remains unchanged, so the final obligation in this case is to prove the effect just given for $e'$ (technically, preceded by $I\rhd$) is a subeffect of $\gamma=\gamma_c\rhd(\gamma_b\rhd\gamma_c)^*$, which relies crucially on iteration properties P2 and P5:
\[\begin{array}{rcl}
\gamma_c\rhd((\gamma_b\rhd\gamma_c\rhd(\gamma_b\rhd\gamma_c)^*)\sqcup I)
&\sqsubseteq&
\gamma_c\rhd(((\gamma_b\rhd\gamma_c)\rhd(\gamma_b\rhd\gamma_c)^*)\sqcup I)\\
&\sqsubseteq&
\gamma_c\rhd(((\gamma_b\rhd\gamma_c)^*)\sqcup I)\\
&\sqsubseteq&
\gamma_c\rhd((\gamma_b\rhd\gamma_c)^*)\\
\end{array}
\]
\end{itemize}
\end{myproof}
\begin{theorem}[Type Preservation]
For all $Q$, $\sigma$, $e$, $e'$, $\Sigma$, $\tau$, $\gamma$, and $\gamma'$, if
$\epsilon;\Sigma\vdash e : \tau \mid \gamma$,
$ Q\vdash \sigma : \Sigma$,
$\delta\le\Sigma$, and
$\sigma,e\overset{\gamma'}{\rightarrow}^*_Q \sigma',e'$,
then there exist $\Sigma'$, $\gamma''$ such that
$\epsilon;\Sigma'\vdash e' : \tau \mid \gamma''$,
$Q\vdash\sigma':\Sigma'$,
$\Sigma\le\Sigma'$, and
$\gamma'\rhd\gamma''\sqsubseteq\gamma$.
\end{theorem}
\ifTR
\begin{myproof}
By straightforward induction on the transitive reduction relation, applying Lemma \ref{lem:onestep_preservation} in the inductive case.
\end{myproof}
\fi

\ifTR
\begin{theorem}[One Step Progress]
For all $\Sigma$, $e$, $\tau$, $\gamma$, $\sigma$, if $\epsilon;\Sigma\vdash e : \tau \mid \gamma$ and $Q\vdash\sigma:\Sigma$, then either $e$ is a value, $e$ is stuck on a primitive application that is not defined, or there exists some $e'$, $\sigma'$, and $\gamma'$ such that $\sigma,e\rightarrow^{\gamma'}_Q \sigma',e'$.
\end{theorem}
\begin{myproof}
Standard, by induction on the typing derivation.
\begin{itemize}
\item Cases \textsc{T-Prim}, \textsc{T-Var}, \textsc{T-Lambda}, \textsc{T-Bool}, \textsc{T-TyLambda}, \textsc{T-Unit}: These are all immediately values.
\item Case \textsc{T-App}: First apply the inductive hypothesis for $e_1$'s typing derivation.  If it is stuck on an undefined primitive reduction, so is this application.  If it reduces, then so does this by the first context reduction for application.  If it is a value, apply the inductive hypothesis for $e_2$, repeating the same reasoning.  In the case that both are values, then by Lemma \ref{lem:canonical} (Canonical Forms), $e_1$ has one of two forms:
	\begin{itemize}
	\item $e_1=(\lambda y\ldotp e_b)$ for some variable and function body. So the application steps to $e_b[e_2/y]$ with the identity effect by \textsc{E-App}.
	\item $e_1=p_i\;\overline{v}$ for some primitive that is not yet fully applied.  Thus the overall term is $p_i\;\overline{v}\;e_2$ (where we know $e_2$ is a value).  If $\llbracket p_i\;\overline{v}\;e_2\rrbracket$ is defined, then the term steps by \textsc{E-Prim}.  Otherwise the overall expression is itself a value (a partial application of a primitive).
	\end{itemize}
\item Case \textsc{T-If}: Similar to \textsc{T-App} in reducing the condition.  When the condition is a value, Lemma \ref{lem:canonical} produces the result that the condition is either \textsc{true} or \textsc{false}, and either \textsc{E-IfTrue} or \textsc{E-IfFalse} applies.
\item Case \textsc{T-While}: While loops are macro-expanded to conditionals by \textsc{E-While} when they are in reduction position.
\item Case \textsc{T-TyApp}: Similar to $\textsc{T-App}$. When the expression being applied is a value, Canonical Forms gives either a partial application of a primitive (in which case the type application either reduces, or is another partial application value), or a type-lambda (in which case substitution occurs via \textsc{E-TApp}).
\end{itemize}
\end{myproof}
\fi

\section{Relationships to Semantic Notions of Effects}
\label{sec:semantics}

Our notion of an effect quantale is motivated by generalizing directly from the form of effect-based type judgments.  In parallel with our work, there has been a line of semantically-oriented work to generalize monadic semantics to capture sequential effect systems (indeed, this is where our use of the term ``sequential effect system'' originates).  Here we compare to several recent developments: Tate's productors (and algebraic presentation as effectoids)~\cite{tate13}, Katsumata's effect-indexed monads~\cite{katsumata14}, and Mycroft, Orchard, and Petricek's joinads (and algebraic presentation in terms of jonoids)~\cite{mycroft16}.

All of this work is done primarily in the setting of category theory, by incrementally considering the categorical semantics of desirable effect combinations (in contrast to our work, working by generalizing actual effect systems).  Fortunately, each piece of work also couples the semantic development with an algebraic structure that yields an appropriate categorical structure, and we can compare directly with those without appealing to much category theory.  None of the following systems consider effect polymorphism or give more than a passing mention of iteration, though given the generality of the technical machinery, we cannot say any of the following are incompatible with these ideas --- only that their use has not been considered.  In contrast, we showed (Section \ref{sec:soundness}) that effect quantales are compatible with these ideas.
Effect domains that depend on program semantics (e.g., singleton effects) have also not been considered in this semantic work, while we consider indexed effect quantales whose effects depend on program values.
Of the three families of semantic work we compare to, only Mycroft et al.\ go so far as to consider conditionals and discuss iteration, which are ignored (in favor of other important issues) in Tate and Katsumata's work.

Overall, Tate and Katsumata's work studies structures which are strict generalizations of effect quantales (i.e., impose fewer constraints than effect quantales), and any effect quantale can be translated directly to Tate's effectoids or Katsumata's partially ordered effect monoid.  Tate and Katsumata demonstrate that their structures are \emph{necessary} to capture certain parts of any sequential effect system --- a powerful general claim.  By contrast, we demonstrate that with just a bit more structure than either of these, effect quantales become \emph{sufficient} to formalize a range of real sequential effect systems.
Mycroft et al.'s work does consider a full programming language, but studies a different set of structures than we do (block-structured parallelism rather than iteration).

\subsection{Productors and Effectoids}
Tate~\cite{tate13} sought to design the maximally general semantic notion of sequential composition, proposing a structure called \emph{productors}, and a corresponding algebraic structure for source-level effects called an \emph{effector}.  Effectors, however, include models of analyses that are not strictly modular (e.g., can special-case certain patterns in source code for more precise effects)~\cite[Section 5]{tate13}.
To model the strictly compositional cases like syntactic type-and-effect systems, he also defines a semi-strict variant called an \emph{effectoid} (using slightly different notation):
\begin{definition}[Effectoid~\cite{tate13}]
An \emph{effectoid} is a set \textsc{Eff} with a unary relation $\mathsf{Base}(-)$, a binary relation $-\le-$, and a ternary relation $-\fcmp-\mapsto-$, satisfying
\begin{itemize}
\item \textsf{Identity}:
$\forall\varepsilon,\varepsilon'\ldotp
	(\exists \varepsilon_\ell\ldotp \mathsf{Base}(\varepsilon_\ell)\wedge \varepsilon_\ell\fcmp\varepsilon\mapsto\varepsilon')
	\Leftrightarrow
	\varepsilon\le\varepsilon'
	\Leftrightarrow
	(\exists \varepsilon_r\ldotp \mathsf{Base}(\varepsilon_r)\wedge \varepsilon\fcmp\varepsilon_r\mapsto\varepsilon')
$
\item \textsf{Associativity}:
$\forall\varepsilon_1,\varepsilon_2\varepsilon_3,\varepsilon\ldotp
	(\exists\overline{\varepsilon}\ldotp \varepsilon_1\fcmp\varepsilon_2\mapsto\overline{\varepsilon}\wedge\overset{\varepsilon}\fcmp\varepsilon_3\mapsto\varepsilon)
	\Leftrightarrow
	(\exists\hat{\varepsilon}\ldotp \varepsilon_2\fcmp\varepsilon_3\mapsto\hat{\varepsilon}\wedge\varepsilon_1\fcmp\hat{\varepsilon}\mapsto\varepsilon)$
\item \textsf{Reflexive Congruence}:
	\begin{itemize}
	\item $\forall\varepsilon\ldotp \varepsilon\le\varepsilon$
	\item $\forall\varepsilon,\varepsilon'\ldotp\mathsf{Base}(\varepsilon)\wedge\varepsilon\le\varepsilon'\Longrightarrow\mathsf{Base}(\varepsilon')$
	\item $\forall\varepsilon_1,\varepsilon_2,\varepsilon,\varepsilon'\ldotp \varepsilon_1\fcmp\varepsilon_2\mapsto\varepsilon\wedge\varepsilon\le\varepsilon'\Longrightarrow\varepsilon_1\fcmp\varepsilon_2\mapsto\varepsilon'$
	\end{itemize}
\end{itemize}
\end{definition}
Intuitively, \textsf{Base} identifies effects that are valid for programs with ``no'' effect --- e.g., pure programs, empty programs.  Tate refers to such effects as \emph{centric}.  The binary relation $\le$ is clearly a partial order for subeffecting, and $-\fcmp-\mapsto-$ is (relational) sequential composition.  The required properties imply that the effectoid's sequential composition can be read as a non-deterministic function producing the minimal composed effect \emph{or any supereffect thereof}, given that the sequential composition relation includes left and right units for any effect, and that \textsf{Base} and the last position of composition respect the partial order on effects.

Given Tate's aim at maximal generality (while retaining enough structure for interesting reasoning about sequential composition), it is perhaps unsurprising that all but the most degenerate effect quantale yields an effectoid by flattening the monoid and semilattice structure into the appropriate relations:
\begin{lemma}[Quantale Effectoids]
For any nontrivial effect quantale $Q$ (one with more elements than $\top$), there exists an effectoid $E$ with the following structure:
\begin{itemize}
\item $\textsc{Eff}=E/\{\top\}$
\item $\mathsf{Base}(a) \overset{def}{=} I \sqsubseteq a\land a\neq\top$
\item $a \le b \overset{def}{=} a \sqsubseteq b \land b\neq\top$
\item $a\fcmp b \mapsto c \overset{def}{=} a\rhd b = c' \wedge c'\sqsubseteq c\land c\neq\top$
\end{itemize}
\end{lemma}
\begin{myproof}
The laws follow almost directly from the effect quantale laws.  In the identity property, both left and right units are always chosen to be $I$.  Associativity follows directly from associativity of $\rhd$ and isotonicity.  The reflexive congruence laws follow directly from the definition (and transitivity) of $\sqsubseteq$.
Note that we removed the top (error) element, representing failure by missing entries in the relations.
\end{myproof}

A bit more surprising, perhaps, is that many effectoids directly yield quantales:
\begin{lemma}
For any effectoid $E$ with a least centric element, and whose underlying partial order is a join semilattice, and which has a least result for any defined sequential composition, there exists an effect quantale $Q$ such that:
\begin{itemize}
\item $E_Q=\textsc{Eff}_E\uplus{\mathsf{Err}}$
\item $\top=\mathsf{Err}$ (a synthetic error element)
\item $\sqcup$ performs the assumed binary join extended for new top element \textsf{Err}.
\item $a\rhd b$ produces the least $c$ such that $a\fcmp b\mapsto c$ when defined, or \textsf{Err} when there is no such $c$
 such that $a\fcmp b\mapsto c$ (by assumption, $a\fcmp b$ is undefined or has a least element).
\item $I$ is assumed the least centric element
\end{itemize}
\end{lemma}
\ifTR
\begin{myproof}
The effect quantale laws follow directly from the effectoid laws and the minor extension of translating undefined compositions into explicit errors.
\end{myproof}
\fi
Tate calls effectoids with a least result for any defined sequential composition \emph{principalled}, and notes that they are common.

Essentially, in the case where the effectoid's partial order corresponds to a join semilattice with a single unit for sequencing and deterministic (modulo subsumption) sequencing, the two notions coincide.  This strongly suggests that our generalization from the type judgments of a few specific effect systems, rather than from semantic notions, did not cost much in the way of generality.  It also clarifies exactly when effectoids are more general: when effects form a partial order but \emph{not} a join semilattice (no unique least upper bound of any pair), have no universal unit for sequencing, or have non-deterministic sequencing results.  We are unaware of any complete source-level type-and-effect system with these properties.

\ifTR
The need for multiple base elements, and for \textsf{Base} and sequential composition to subsume subeffecting on their last arguments, arises from the category-theoretic approach.  There, type \emph{derivations} (rather than simply programs) are translated to morphisms in a category, where the same program expression translated under different contexts will produce different (but related) morphisms.  The basic cateogory laws require morphism composition to be associative, so translating a derivation that intuitively applies some effect operation with least effects followed by a general subeffecting judgment (Tate's example calculus includes a subeffecting rule) would be considered equal to a judgment that directly produces a super-effect of a ``smaller'' effect.  This is an important part of ensuring the categorical semantics are sensible, but is akin to the standard choice in type systems with subtyping of including a general subtyping rule or applying the subtyping relation throughout the type system: two choices that are equivalent with regards to soundness.
\fi

\subsection{Effect-indexed Monads, a.k.a.\ Graded Monads}
Katsumata~\cite{katsumata14} pursues an independent notion of general sequential composition, where effects are formalized semantically as a form of type refining monad: a $T\;e\;\sigma$ is a monadic computation producing an element of type $\sigma$, whose effect is bounded by $e$ (which classifies a subset of such computations).  Based on general observations, Katsumata speculates that sequential effects form at least a preordered monoid, and goes on to validate this (among other interesting results related to the notion of effects as refinements of computations).  Katsumata shows categorically that these \emph{effect indexed monads} (which later came to be known as \emph{graded monads} to avoid confusion with other forms of indexing) are also a specialization of Tate's productors, exactly when the productor is induced by an effectoid derived from a partially-ordered monoid.  Our notion of effect quantales directly induces a partially ordered monoid $(E,\sqcup,\rhd,I)$ satisfying the appropriate laws.
However, the effectoid equivalent to this translation is not quite the same as the direct effectoid described earlier: graded monads (particularly the po-monoids) do not directly model partiality, while effectoids can.
Setting this minor discrepancy aside (e.g., one could impose type system restrictions on its use, as we did in our type system)
the relaxation between effect quantales and graded monads is due to relaxing the bounded join semilattice to a partial order, and the change from graded monads to effectoids (and thus productors) is due primarily to relaxing the rules for sequencing identity and determinism of sequencing.  Katsumata does note briefly that many interesting effect systems rely on join-semilattices, but does not explore this specific class of graded monads in depth.

\subsection{Joinads and Joinoids}
\label{sec:joinoids}
Mycroft, Orchard, and Petricek~\cite{mycroft16} further extend graded monads to \emph{graded conditional joinads}, and similar to Tate, give a class of algebraic structures --- joinoids --- that give rise to their semantic structures.  As their base, they take graded monads, further assume a ternary conditional operator $?{}:(-,-,-)$ modeling conditionals whose branch approximation may depend on the conditional expression's effect, and parallel composition $\&$ suitable for fork-join style concurrency.

Their ternary operator is motivated by considerations of sophisticated effects such as control effects like backtracking (e.g., continuations).  From their ternary operator, they derive a binary join, and therefore a partial order.  However, their required laws for the ternary operator include only a right distributivity law because effects from the conditional expression itself do not in general distribute into the branches.  Thus their derived semilattice structure satisfies only the right distributivity law $(a\sqcup b)\rhd c = (a\rhd c)\sqcup(b\rhd c)$, and not, in general, the left-sided equivalent.  They also do not require ``commutativity'' of the branch arguments. This means that joinoids, in general, do not give rise to effect quantales --- some (small) amount of structure is not necessarily present --- and that in general they validate fewer equivalences between effects.
An effect quantale can induce a ternary operator that ignores its first argument by simply taking the join of its other arguments, in which case Mycroft et al.'s derived partial order coincides with that derived from the quantale's join.  As with the relation to graded monads this translates error element concretely rather than directly modeling partiality.

Joinads originally arose as an extension to monads that captures a class of combinators typical of composing parallel and concurrent programs in Haskell, in particular a \emph{join} (unrelated to lattices) operator of type $M\;A\rightarrow M\;B\rightarrow M\;(A\times B)$.  This is a natural model of fork-join-style parallel execution, and gives rise to the $\&$ operator of joinoids, which appears appropriate to model the corresponding notion in systems like Nielson and Nielson's effect system for CML communication behaviors~\cite{nielson1993cml}, which is beyond the space of operations considered for effect quantales.  However, $\&$ is inadequate for modeling the unstructured parallelism (i.e., explicit thread creation and termination, or task-based parallelism) found in most concurrent programming languages, so we did not consider such composition when deriving effect quantales.  We would like to eventually extend effect quantales for unstructured concurrent programming: this is likely to include adapting ideas from concurrent program logics that join asynchronously~\cite{dodds09}, but any adequate solution should be able to induce an operation satisfying the requirements of joinoids' parallel composition.

Ultimately, any effect quantale gives rise to a joinoid, by using the effect quantale's join for both parallel composition and to induce the ternary operator outlined above.

\paragraph*{Fixed Points}
Mycroft et al.\ also give brief consideration to providing iteration operators through the existence of fixed points, noting the possibility of adding one type of fixed point categorically, which carried the undesirable side effect of requiring sequential composition to be idempotent: $\forall b\ldotp b\rhd b = b$.  This is clearly too strict, and prohibits equivalents of both the lockset and atomicity effect quantales we studied.  They take this as an indication that every operation should be explicitly provided by an algebra, rather than attempting to derive operators.
By contrast, our closure operator approach not only imposes semantics that are by construction compatible with a given sequential composition operator, but critically coincide with manual definitions for existing systems.

\subsection{Limitations of Semantics-Based Work}
The semantic work on general models of sequential effect systems has not seriously addressed iteration.  As discussed above, Mycroft et al.\ note that a general fixed point map could be added, but this forces $a\rhd a=a$ for all effects $a$, which is too restrictive to model the examples we have considered.
Our approach to inducing an iteration operation through closure operators on posets should be generalizable to each of the semantic approaches we discussed.  The semantics of such an approach are, broadly, well-understood, as closure operators on a poset are equivalent to a certain monad on a poset category; note that the three properties of closure operators --- extensiveness, idemotence, and monotonicity --- correspond directly to the formulation of a monad in terms of return, join (a flattening operation $M(M\;A)\rightarrow M\;A$ unrelated to lattices or joinoids),
and fmap.  
The semantic work discussed also omits treatment of polymorphism, and singleton or dependent types.  As a result, their claim of adequacy for sequential effect systems is limited, whereas we have provide in Section \ref{sec:modeling2} a direct implementation of a non-trivial composite sequential effect system in terms of effect quantales.
On the other hand, their claims to generality are much stronger than ours, not only because the corresponding algebraic structures are less restrictive, but because they derived these structures by focusing on a few key elements common to all sequential effect systems (aside from the parallel combination studied for joinads) rather than directly attempting to generalize from concrete examples of sequential effect systems.  Ultimately we view our work as strictly complementary to this categorical work --- the latter is foundational and deeply general, while ours is driven by practice of sequential effect systems.
Our work fills in a missing connection between these approaches and the concrete syntactic sequential effect systems most have studied.

The categorical semantics of polymorphism and dependent types (including singleton indexing as we have) are generally well-understood~\cite{phoa92,dybjer1995internal,jacobs1999categorical} and have even gained significant new tools of late~\cite{birkedal2013intensional}, so the work discussed here should be compatible with those ideas, even if it requires adjustment.
However, these related approaches would also need to be extended to account for substitution into effects that may mention program values; the notion of collapsibility will require an analogue in semantic accounts.

\section{Modeling Prior Effect Systems in a Generic Framework}
\label{sec:modeling2}
This section demonstrates that we can model significant prior type systems by embedding into our core language.  Embedding here means a type-and-effect-preserving, but not necessarily semantics-preserving translation.  Our langauge is generic, but clearly lacks concurrency, exception handling, and other concrete computational effects.  Instead, we show how to model relevant primitives in our core language, giving derived type rules for those constructs, and translate type judgements to prove we would at least accept the same programs.

\subsection{Types for Safe Locking and Atomicity}
\begin{figure}\scriptsize
\begin{mathpar}
\fbox{$\Gamma\vdash e : a$}
\quad
\inferrule[\scriptsize EXP CONST]{ }{\Gamma\vdash c : B}
\quad
\inferrule[\scriptsize EXP LOC]{ }{\Gamma\vdash m : B}
\quad
\inferrule[\scriptsize EXP FUN]{\Gamma\vdash e : \Gamma(f)}{\Gamma\vdash f(\overline{x}) e : B}
\quad
\inferrule[\scriptsize EXP PRIM]{\Gamma\vdash e_i : a_i}{\Gamma\vdash p(\overline{e}) : (a_1;\ldots;a_n;\Gamma(p))}
\quad
\inferrule[\scriptsize EXP READ]{ }{\Gamma\vdash x_\epsilon : B}
\quad
\inferrule[\scriptsize EXP RRACE]{ }{\Gamma\vdash x_\bullet : A}
\and
\inferrule[\scriptsize EXP ASSIGN]{\Gamma\vdash e : a}{\Gamma\vdash x_\epsilon := e : (a;B)}
\quad
\inferrule[\scriptsize EXP RASSIGN]{\Gamma\vdash e : a}{\Gamma\vdash x_\bullet := e : (a;A)}
\quad
\inferrule[\scriptsize EXP LET]{\Gamma\vdash e_1 : a_1 \\ \Gamma\vdash e_2:a_2}{\Gamma\vdash\mathsf{let}\;x=e_1\;\mathsf{in}\;e_2 : (a_1;a_2)}
\quad
\inferrule[\scriptsize EXP IF]{\Gamma\vdash e:a \\ \Gamma\vdash e_i:b:i}{\Gamma\vdash\mathsf{if}\;e\;e_1\;e_2 : (a;(b_1\sqcup b_2))}
\and
\inferrule[\scriptsize EXP WHILE]{\Gamma\vdash e_1:a_1\\\Gamma\vdash e_2:a_2}{\Gamma\vdash\mathsf{while}\;e_1\;e_2 : (a_1;(a_2;a_1)^*)}
\quad
\inferrule[\scriptsize EXP INVOKE]{\Gamma\vdash e:a \\ \Gamma\vdash e_i:a_i}{\Gamma\vdash e^F(\overline{e}) : (a_1;\ldots;a_n;(\sqcup_{f\in F}\Gamma(f)))}
\quad
\inferrule[\scriptsize EXP FORK]{\Gamma\vdash e:a}{\Gamma\vdash\mathsf{fork}\;e : A}
\quad
\inferrule[\scriptsize EXP ATOMIC]{\Gamma\vdash e:a \\ a\sqsubseteq A}{\Gamma\vdash\mathsf{atomic}\;e : a}
\end{mathpar}
\vspace{-0.5cm}
\caption{Flanagan and Qadeer's type and effect system for atomicity of \textsc{CAT} programs.}
\label{fig:flanagan}
\end{figure}

Here we briefly recall the details of Flanagan and Qadeeer's earlier work on a type system for atomicity~\cite{flanagan2003tldi} (the full version~\cite{flanagan2003atomicity} requires substantially more space and extends Java --- modeling objects would require a more sophisticated type system for embedding).
Flanagan and Qadeer's \textsc{CAT} language (Figure \ref{fig:flanagan}) is minimalist, defined in terms of a family of primitives (like our core language), with named functions, racing and race-free heap accesses, expected control constructs, and atomic blocks (which must be atomic).
They use semicolons for sequencing of atomicity effects.
For maximal minimalism, they assume some \emph{other} type system has already analyzed the program and identified which heap accesses are racy and which are well-synchronized.  For completeness, we will embed into an instantiation of our framework that itself distinguishes well-synchronized and racy reads, and establish conditions under which their abstract notion of well-synchronized is compatible.  Thus this section develops a hybrid of Flanagan and Abadi's \emph{Types for Safe Locking}~\cite{safelocking99} and Flanagan and Qadeer's \emph{Types for Atomicity}~\cite{flanagan2003tldi}, further extended to track locks in a flow-sensitive manner (the former uses \texttt{synchronized} blocks, the latter does not track locks itself).
Recall that in the former, a concurrent functional language with heap is extended by locks, and the reference type is indexed by a singleton lock identity.  The type system tracks the set of locks held at each program point (there, scoped by lexically scoped \textsf{synchronized} blocks), and ensures that any access to a heap location guarded by some lock occurs while that lock is held.  This forms the foundation of the ideas behind the better-known \textsc{RCC/Java}~\cite{rccjava00}, which extends these ideas to the full Java language.  We add additional read and write primitives that may race, to model the atomicity work.

\begin{figure}\scriptsize
\[
\begin{array}{l}
Q(X) = \mathcal{L}(X)\otimes\mathcal{A}\\
M \in \mathsf{LockNames}\rightharpoonup\mathsf{Bool}\\
H \in \mathsf{Location}\rightharpoonup\mathsf{Term}\\
\mathsf{State} = M\times H\\
K(\mathsf{lock}) = \star\\
K(\mathsf{ref}) = \star\Rightarrow\star\Rightarrow\star\\
\\
\inferrule{
    \forall l\in\mathsf{dom}(m)\ldotp \Sigma(l)=\mathsf{lock}\\\\
    \forall r\in\mathsf{dom}(h)\ldotp \epsilon;\Sigma\vdash h(r):\Sigma(r)\mid I
}{
    Q\vdash (m,h) : \Sigma
}
\end{array}
\begin{array}{rcl}
\delta(\mathsf{new\_lock}) &=& \mathsf{unit}\overset{B}{\rightarrow}\mathsf{lock}\\
\delta(\mathsf{acquire}) &=& \Pi x:\mathsf{lock}\xrightarrow{(\emptyset,\{x\})\otimes R}\mathsf{unit}\\
\delta(\mathsf{release}) &=& \Pi x:\mathsf{lock}\xrightarrow{(\{x\},\emptyset)\otimes L}\mathsf{unit}\\
\delta(\mathsf{alloc}) &=& \Pi x:\mathsf{lock}\overset{B}{\rightarrow}\forall \alpha::\star\overset{B}{\rightarrow} \tau \overset{B}{\rightarrow}\mathsf{ref}\;S(x)\;\tau\\
\delta(\mathsf{read_\bullet}) &=& \Pi x:\mathsf{lock}\overset{B}{\rightarrow}\forall \alpha::\star\overset{B}{\rightarrow} \mathsf{ref}\;S(x)\;\tau {\xrightarrow{(\emptyset,\emptyset)\otimes A}} \tau\\
\delta(\mathsf{read_\epsilon}) &=& \Pi x:\mathsf{lock}\overset{B}{\rightarrow}\forall \alpha::\star\overset{B}{\rightarrow} \mathsf{ref}\;S(x)\;\tau {\xrightarrow{(\{x\},\{x\})\otimes B}} \tau\\
\delta(\mathsf{write_\bullet}) &=& \Pi x:\mathsf{lock}\overset{B}{\rightarrow}\forall \alpha::\star\overset{B}{\rightarrow} \mathsf{ref}\;S(x)\;\tau \overset{B}{\rightarrow}\tau{\xrightarrow{(\emptyset,\emptyset)\otimes A}} \tau\\
\delta(\mathsf{write_\epsilon}) &=& \Pi x:\mathsf{lock}\overset{B}{\rightarrow}\forall \alpha::\star\overset{B}{\rightarrow} \mathsf{ref}\;S(x)\;\tau \overset{B}{\rightarrow}\tau{\xrightarrow{(\{x\},\{x\})\otimes B}} \tau\\
\delta(\mathsf{req\_atomic}) &=& (\mathsf{bool}\overset{A}{\rightarrow}\mathsf{unit})\overset{B}{\rightarrow}\mathsf{unit}
\end{array}
\]
\vspace{-2em}
\[
\begin{array}{l}
\llbracket\mathsf{new\_lock}\;\_\rrbracket((m,h))(\Sigma) = l,(m[l\mapsto\mathsf{false}],h),\Sigma[l\mapsto\mathsf{lock}]~\textrm{for next $l\not\in\mathsf{dom}(m)$}\\
\llbracket\mathsf{acquire}\;l\rrbracket((m[l\mapsto\mathsf{false}],h))(\Sigma) = (),(m[l\mapsto\mathsf{true}],h),\Sigma\\
\llbracket\mathsf{release}\rrbracket((m[l\mapsto\mathsf{true}],h))(\Sigma) = (),(m[l\mapsto\mathsf{false}],h),\Sigma\\
\llbracket\mathsf{alloc}\;l\;\tau\;v\rrbracket((m,h))(\Sigma) = \ell,(m,h[\ell\mapsto v),\Sigma[\ell\mapsto\mathsf{ref}\;S(l)\;\tau]~\textrm{for $\ell\not\in\mathsf{dom}(h)$}\\
\llbracket\mathsf{read_\bullet\;l\;\tau\;\ell}\rrbracket((m,h))(\Sigma) = h(\ell),(m,h),\Sigma\\
\llbracket\mathsf{read_\epsilon\;l\;\tau\;\ell}\rrbracket((m,h))(\Sigma) = h(\ell),(m,h),\Sigma\\
\llbracket\mathsf{write_\bullet\;l\;\tau\;\ell\;v}\rrbracket((m,h))(\Sigma) = v,(m,h[\ell\mapsto v]),\Sigma\\
\llbracket\mathsf{write_\epsilon\;l\;\tau\;\ell\;v}\rrbracket((m,h))(\Sigma) = v,(m,h[\ell\mapsto v]),\Sigma\\
\llbracket\mathsf{req\_atomic}\;f\rrbracket((m,h))(\Sigma)=(),(m,h),\Sigma
\end{array}
\]
\vspace{-1em}
\caption{Parameters to model Flanagan and Abadi's \emph{Types for Safe Locking}~\cite{safelocking99} (a sequential variant) and Flanagan and Qadeer's \emph{Types for Atomicity}~\cite{flanagan2003tldi} in our framework.  We sometimes omit the locking component of effects when it is simply $(\emptyset,\emptyset)$ to improve readability.}
\label{fig:locking_params}
\end{figure}

We define in Figure \ref{fig:locking_params} the parameters to the language framework needed to model locks, mutable heap locations, and lock-indexed reference types, and the primitives to manipulate them.
We define $T_i$ by giving $K$ (which is defined over $T_i$), and define $p_i$ as $\mathsf{LockNames}\uplus\mathsf{Location}\uplus\mathsf{dom}(\delta)$ (locks, heap locations, and primitive operations).
The state consists of a lock heap, mapping locks to a boolean indicating whether each lock is held, and a standard mutable store.  The reference type is indexed by a lock (lifted to a singleton type).
Primitives include lock allocation; lock acquisition and release primitives whose effects indicate both the change in lock claims and the mover type;
allocation of data guarded by a particular lock; racing ($\bullet$) and well-synchronized ($\epsilon$) reads and writes, with effects requiring (or not) lock ownership as appropriate; and one further primitive for requiring atomicity.
We give a stylized definition of the (partial) semantics function for primitives as acting on not only states but also state types, giving the monotonically increasing state type for each primitive, as required of the parameters.  We also omit restating the dynamic effect in our $\llbracket-\rrbracket$; we take it to be the final effect of the corresponding entry in $\delta$ with appropriate value substitutions made --- as required by the type system.
The definitions easily satisfy the primitive preservation property assumed by the type system.
We take as the partial order on $\textsf{StateEnv}$ the standard partial order on partial functions, with $\delta$ as its least element.

These parameters are adequate to write and type terms like the following atomic function that reads from a supplied lock-protected reference (permitting syntactic sugar for brevity):
\[
\begin{array}{@{\qquad}l}
\emptyset\vdash \lambda x\ldotp\lambda r\ldotp \mathsf{acquire}\;x; \mathsf{let}\;y=\mathsf{read}_\epsilon\;x\;[\mathsf{bool}]\;r\;\mathsf{in}\;(\mathsf{release}\;x; y)\\
\qquad : \left(\Pi x:\mathsf{lock}\xrightarrow{(\emptyset,\emptyset)\otimes B}
	\Pi r:\mathsf{ref}\;\mathcal{S}(x)\;\mathsf{bool}\xrightarrow{(\emptyset,\emptyset)\otimes A}\mathsf{bool}\right) \mid (\emptyset,\emptyset)\otimes B
\end{array}
\]

\textsc{CAT} is a property multi-threaded language, while our language is not.  As we noted earlier, our aim is to preserve well-typing, not dynamic semantics, so our translation of \textsf{fork} will not model concurrent semantics.
Blocks of code that do not fork or rely on other threads should run as expected, though we do not prove this.

\textsc{CAT}'s constants, primitives (\texttt{new\_lock}, etc.), and mutexes can be translated in almost the obvious way for our framework, currying their primitives and extending that set with constants and the mutex names described above.
The tricky bit is that \textsc{CAT} presumes some unspecified race freedom analysis and unspecified type system have already been applied to distinguish racing and well-synchronized reads, and to rule out basic type errors.  Our terms require lock and type information to be explicitly present in the term, so we assume, beyond those unspecified analyses, operations \textsf{LockFor}, \textsf{RefTypeOf}, and \textsf{TypeOf} to extract the relevant local lock names and types.  For a term produced using these operations to type-check in our core language will naturally require a degree of consistency between the unspecified analyses and the checks of our core language for the lock multiset quantale.  However the details are not necessary to work out, because our relation is conditioned on the assumption that the translation does type check in our core language.

Conditionals and while loops are translated in the obvious inductive way --- note that aside from \textsc{CAT}'s type system lacking basic types, the handling of atomicity effects is structured exactly as our rules for those constructs.
To handle currying, we adopt the notations
$\lambda \overline{x}\ldotp e \equiv \lambda x_1\ldots\lambda x_n\ldotp e$ for an $n$-ary closure, and
$e\;\overline{e'} \equiv (\ldots(e\;e_1')\ldots e_n')$ for $n$-ary function application.
Note that when typing the expanded forms, the effects of all but the innermost expanded lambda expression can simply be $I$, making the overall effect of the expanded application the left-to-right sequenced effects of the function and each argument followed by the effect of the inner-most closure.
We also use the shorthand $\mathsf{wraplock}\;e\equiv \mathsf{let}\;x=\mathsf{new\_lock}()\;\mathsf{in}\;(\mathsf{acquire}\;x; \;e;\;\mathsf{release}\;x;())$.  The atomicity of this expression is $A$ if and only if $e$'s atomicity is less than $A$.
Other translations are as follows, omitting analagous primitive translations:
\[
\begin{array}{@{\llbracket}c@{\rrbracket =}l@{\quad}@{\llbracket}c@{\rrbracket =}l}
p(\overline{e}) & p\;\overline{\llbracket e\rrbracket} &
e^F(\overline{e}) & \llbracket e\rrbracket\;\overline{\llbracket e\rrbracket}\\
f(\overline{x})e & (\lambda\overline{x}\ldotp \llbracket e\rrbracket) &
\texttt{atomic}\;e & \mathsf{req\_atomic}\;(\lambda \_\ldotp \mathsf{wraplock}\;\llbracket e\rrbracket);\;\llbracket e\rrbracket \\
\texttt{fork}\;e & \mathsf{let}\;\_=(\lambda\_\ldotp\llbracket e\rrbracket)\;\mathsf{in}\;\mathsf{wraplock}\;() &
    x_\bullet & \mathsf{read}_\bullet\;\langle\mathsf{LockFor}(x)\rangle\;\langle\mathsf{RefTypeOf}(x)\rangle\;x
\end{array}
\]
We assume the translation process produces a mapping from generated subterms back to the original \textsc{CAT} term (specifically, mapping closures back to \textsc{CAT}'s named functions).
\texttt{atomic} expression are translated to capture the expression in a dynamically-meaningless thunk passed as a parameter requiring an atomic effect (recall function bodies can have lesser effects than their types suggest), but run unconditionally.  The unconditional execution allows the actual atomicity of $e$ to be used later, as in \textsc{CAT}.
\texttt{fork} operations are translated in a way that makes the forked thread computationally irrelevant (but, by induction, preserves typeability and effects) and locally carries an atomic effect (by subeffecting from the $B$ to $A$) as in the type rule.

The theorem we would like to prove is that translating any well-typed \textsc{CAT} term produces a term in our core language with the corresponding type and effect.  Unfortunately, \textsc{CAT} is untyped aside from atomicities, so there is no type to translate, and \textsc{CAT} itself cannot check correct use of well-synchronized vs.\ racy reads.  Instead, we prove an ``un-embedding'' lemma by induction on the \textsc{CAT} term:
\begin{lemma}[Unembedding \textsc{CAT} from $\mathcal{L}\otimes\mathcal{A}$]
Given a \textsc{CAT} term $t$, for any $\Gamma$, $\tau$, and effect $l\otimes e\in(\mathcal{L}\otimes\mathcal{A})(\Gamma)$ such that
$\Gamma\vdash \llbracket t\rrbracket : \tau \mid l\otimes e$, under the \textsc{CAT} environment $\hat\Gamma$ mapping each function name to the final effect of
its $n$-ary closure translation, $\hat\Gamma\vdash t : e$.
\end{lemma}
\ifTR
\begin{proof}
By induction on $t$.
\begin{itemize}
\item Case $t=c$: $c$ embeds to a primitive, or a constant boolean or unit.  In either case, because $I=B$, the term is well-typed by \textsc{EXP CONST}.
\item Case $t=m$: This embeds to a primitive location, which will have the appropriate unit effect by \textsc{EXP SYNCLOC}.
\item Case $t=f(\overline{x})e$: This embeds to an n-ary lambda $(\lambda\overline{x}\ldotp \llbracket e\rrbracket)$, which will always have effect $I=B$.
\item Case $t=p(\overline{e})$: This case embeds application of primitives to application of primitives.   By (repeated) inversion on the core language application, each $e_i\in\overline{e}$ is well-typed with effect $a_i$, and the overall effect of the application is the fold of $\rhd$ over the sequence of subexpression effects and the final effect of the primitive.  This corresponds (after simplifying away some unit effects) to the effect $(a_1;\ldots;a_n;\Gamma(p))$ in \textsc{CAT}'s \textsf{EXP PRIM}.
\item Case $t=x_\epsilon$: This embeds to a full application of $\mathsf{read}_\epsilon$, with effect $B$ as required by \textsc{EXP READ}.
\item Case $t=x_\bullet$:  This embeds to a full application of $\mathsf{read}_\bullet$, with effect $A$ as required by \textsc{EXP RRACE}.
\item Case $t=x_\epsilon := e$: This embeds to a full application of $\mathsf{write}_\epsilon$.  By inductive hypothesis, $e$ embeds to a term with atomicity $a$ and can be typed by the same effect in \textsc{CAT}'s effect system.  By (repeated) inversion on the core language typing judgement for the write primitive, the overall effect of the embedded result will be $a\rhd B$ (or, $a;B$).  Using the inductive result for $e$'s effect, \textsc{CAT} also gives the overall write that effect via \textsc{EXP ASSIGN}.
\item Case $t=x_\bullet := e$: Similar to the previous case, but using $\mathsf{write}_\bullet$ and \textsc{EXP RASSIGN}.
\item Case $t=\mathsf{let}\;x=e_1\;\mathsf{in}\;e_2$: Straightforward repeated use of the inductive hypothesis (recall that \textsc{CAT}'s type system only maps function names to atomicities).
\item Case $t=\mathsf{if}\;e\;e_1\;e_2$: Follows from multiple uses of the inductive hypothesis.
\item Case $t=\mathsf{while}\;e_1\;e_2$: Follows from multiple uses of the inductive hypothesis.
\item Case $t=e^F(\overline{e})$: In \textsc{CAT}, no proper type system tracks the latent effect of first-class functions, so \textsc{CAT} assumes that $F$ contains the static name of every closure that may be produced as the result of $e$.  Because we know the translation of this application is well-typed, we know $e$ will have a function type whose latent effect $a$ is an upper bound on the latent effect of the invoked closure.  For $F$ to be accurate, this implies $a\sqsubseteq \sqcup_{f\in F}\hat{\Gamma}(f)$.  The remainder of this case --- producing a valid use of \textsc{EXP INVOKE} --- parallels the bulk of the primitive case.
\item Case $t=\mathsf{fork}\;e$: The valid typing of this translation ensures $e$ is well-typed with some atomicity $a$, but --- because we ensure only type preservation in the absence of proper concurrency --- the translation discards a closure containing $\llbracket e\rrbracket$ without invoking it.
The use of $\mathsf{wraplock}\;()$ ensures the occurrence of an atomic action (forcing atomicity of the fork translation), while the ``throw-away'' of the new thread body's encoding ensures the body is well-typed with \emph{some} valid atomicity.
\item Case $t=\mathsf{atomic}\;e$: A use of \textsf{wraplock} (recall, this is a syntactic shorthand) is atomic ($A$) if and only if the wrapped expression's atomicity is a sub-atomicity of $A$.  Passing a use of this shorthand in a closure to a primitive requiring an atomic body enforces that the body's atomicity must in fact be (a subeffect of) atomic.  This is then discarded without invoking the closure, and a duplicate translation of $e$ is evaluated directly to permit $e$'s atomicity to be used as the atomicity of the overall translation (the ``throw-away'' closure's usage contributes $B$, where $B\rhd A=A$).
\end{itemize}
\end{proof}
\fi

\section{Related and Future Work}
\label{sec:relwork}
The closely related work is split among three major groups: generic effect systems, algebraic models of sequential computation, and concrete effect systems.

\subsection{Generic Effect Systems}
We know of only three generic characterizations of effect systems prior to ours, none of which handles sequential effects or is extensible with new primitives.

Marino and Millstein give a generic model of a static commutative effect system~\cite{marino09} for a simple extension of the lambda calculus.  Their formulation is motivated explicitly by the view of effects as capabilities, which pervades their formalism --- effects there are sets of capabilities, values can be tagged with sets of capabilities, and subeffecting follows from set inclusion.  They do not consider polymorphism (beyond the naive exponential-cost approach of substituting let bindings at type checking).  They do however also parameterize their development by an insightful choice of \emph{adjust} to change the capabilities available within some evaluation context and \emph{check} to check the capabilities required by some redex against those available, allowing great flexibility in how effects are managed.

Henglein et al.~\cite{henglein2005effect} give a simple expository effect system to introduce the technical machinery added to a standard typing judgment in order to track (commutative) effects.  Like like Marino and Millstein they use qualifiers as a primitive to introduce effects.  Because their goals were instructional rather than technical, the calculus is not used for much (it precedes a full typed region calculus~\cite{talpin1992polymorphic}).

Rytz et al.~\cite{rytz12} offer a collection of insights for building managable effect systems, notably the relative effect polymorphism mentioned earlier~\cite{rytz12a} (inspired by anchored exceptions~\cite{vanDooren2005}) and an approach for managing the simultaneous use of multiple effect systems with modest annotation burden.  The system was given abstractly, with respect to a lattice of effects.  Toro and Tanter later implemented this as as a polymorphic extension~\cite{toro2015customizable} to Schwerter et al.'s gradual effect systems~\cite{BanadosSchwerter2014gradual}.  Their implementation is again parameterized with respect to an effect lattice, supporting only closed effects (i.e., no singletons).

\subsection{Algebraic Approaches to Computation}
Our effect quantales are an example of an algebraic approach to modeling sequential computation.  There are many closely-related approaches beyond those discussed in Section \ref{sec:semantics}, such as action logic~\cite{pratt1990action} and
Kleene Algebras (\textsc{KA}s), and Kleene Algebras with Tests (\textsc{KAT}s)~\cite{kozen1997kleene}.  Each of these has some partial order, and an associative binary operation that distributes over joins (and meets).  Some \textsc{KA}s also look very much like effect quantales: one standard example is a \textsc{KA} of execution traces, similar to the effect systems mentioned in Section \ref{sec:other_examples}.
However, Kleene Algebras and relatives are intended to model the semantics of a \emph{possibly-failing} computation, rather than a classification of ``successful'' computations, and thus carries a ring structure unsuitable for effect systems.
The requirement that the \textsc{KA} element $0$ of the partial order is nilpotent for sequencing ($0\cdot x = 0 = x\cdot 0$) but also least in the partial order ($0+x=x=x+0$) makes these systems unsuitable for effect systems.
Some effects have no sensible least element: for locking, this would be an effect $e$ that is considered to both preserve lock sets ($e \sqsubseteq (\emptyset,\emptyset)$) and also change them (e.g., $e\sqsubseteq(\emptyset,\{\ell\})$ among others).  For those systems where a least element does make sense (atomicity without locking, or subsuming commutative effects), their least element $\bot$ is always the identity for sequencing --- $\bot\rhd x = x = x\rhd\bot$.  The ring requirements would require $A\rhd B\rhd A=B$ for atomicity, which fails to reflect that such a sequence is not atomic.

\subsection{Concrete Effect Systems}
\label{sec:concrete}
We discussed several example sequential effect systems throughout, notably Flanagan and Abadi's \emph{Types for Safe Locking}~\cite{safelocking99} (the precursor to \textsc{RCC/Java}~\cite{rccjava00}), and Flanagan and Qadeer's \emph{Types for Atomicity}~\cite{flanagan2003tldi} (again a precursor to a full Java version~\cite{flanagan2003atomicity}).  This atomicity work is one of the best-known examples of a sequential effect system.  Coupling the atomicity structures developed there with a sequantial version of lockset tracking for unstructured locking primitives gives rise to interesting effect quantales, which can be separately specified and then combined to yield a complete effect system.

Suenaga gives a sequential effect system for ensuring deadlock freedom in a language with unstructured locking primitives~\cite{suenaga2008type}, which is the closest example we know of to our lockset effect quantale.  However, Suenaga's lock tracking is structured a bit differently from ours: he tracks the state of a lock as either explicitly present but unowned (by the current thread), or owned by the current thread, thus not reasoning about recursive lock acquisition.  This is isomorphic to a \emph{set}, rather than a multiset, of locks (a subset of a known set of all locks), and thus checks a different property than our lockset quantale.
In fact, most prior type systems tracking owned locks treat only this binary property.  This discrepancy between prior work and our lockset quantales leads to interesting, and slightly surprising subtleties.

Our first attempt to define the locking effect quantale sought to use only \emph{sets} of locks, rather than \emph{multisets}, and to prohibit recursive lock acquisition.  Indeed, such an effect quantale can be defined, satisfying all required properties, for a fixed set of locks.
But once the set of locks is a parameter, the resulting indexed effect quantale is not collapsible!
Viewing this in terms of the type system, consider the term $f=(\lambda l_1\ldotp\lambda l_2\ldotp \mathsf{acquire}\;l_1;\mathsf{acquire}\;l_2)$, which would have type $\Pi l_1:\mathsf{lock}\overset{I}{\rightarrow}\Pi l_2:\mathsf{lock}\overset{(\emptyset,\{l_1,l_2\})}{\rightarrow}\mathsf{unit}$ (ignoring atomicity).
Intuitively, applying this function to the same lock $x$ twice ($f\;x\;x$) would eventually substitute the same value for $l_1$ and $l_2$, yielding an expected overall effect of $(\emptyset,\{x\})$ --- the number of locks acquired shrank because the set would collapse, though the underlying term would try to acquire the same lock twice.  Moreover, after reducing the second application, the resulting term would no longer by type-correct, as $(\emptyset,\{x\})\rhd(\emptyset,\{x\})=\mathsf{Err}$ when holding a lock twice cannot be represented!  This is why the \emph{set}-based lock tracking is not collapsible.  Using multisets as we do in Section \ref{sec:modeling} fixes this problem.
Suenaga does not encounter this, because his lack of closures and linear lock ownership do not permit two variables used for locking to later be unified by substitution.  Other work such as \textsc{RCC/Java}~\cite{rccjava00} avoids the issue because while the system uses sets, the dynamic semantics permit recursive acquisition and count recursive claims in the evaluation contexts.

Many other systems that are not typically presented as effect systems can be modeled as sequential
effect systems.  Notably this includes systems with flow-sensitive additional contexts (e.g., sets
of capabilities) as alluded to in Section \ref{sec:bg}, or fragments of type information in systems
that as-presented perform strong updates on the local variable contexts (e.g., the state transitions
tracked by typestate~\cite{WolffGTA11,Garcia2014FTP}, though richer systems require dynamic reflection
of typestate checks into types~\cite{Sunshine2011}, which is a richer form of dependent effects than our framework currently tracks).  Other forms of behavioral type systems have at least a close correspondence to known effect systems, which are likely to be adaptable to our framework in the future: consider the similarity between session types~\cite{Honda2008} and Nielson and Nielson's effect system for communication in CML~\cite{nielson1993cml}.

\subsection{Limitations and Future Work}
There remain a few important aspects of sequential effect systems that neither we, nor related work on semantic characterizations of sequential effects, have considered.
One important example is the presence of a masking construct~\cite{lucassen88,gifford86} that
locally supresses some effect, such as try-catch blocks or \texttt{letregion} in region calculi.
Another is serious consideration of control effects, which are alluded to in Mycroft et al.'s work~\cite{mycroft16}, but otherwise have not been directly considered in the algebraic characterizations of sequential effects.

Our generic language carries some additional limitations.  
It lacks subtyping and ``subeffecting,'' which enhance usability of the system, but these should not present any new technical difficulties.  
It also lacks support for adding new evaluation contexts through the parameters, which is important
for modeling constructs like \texttt{letregion}.
Allowing this would require more sophisticated machinery for composing partial semantic
definitions~\cite{birkedal2013intensional,Delaware2013MLC,Delaware2013MMM}.

Beyond the effect-flavored variation~\cite{lucassen88,talpin1992polymorphic} of parametric polymorphism and the polymorphism arising from singleton types as we consider here, the literature contains bounded~\cite{Grossman2002Cyclone} (or more generally, constraint-based) effect polymorphism, and unique ``lightweight'' forms of effect polymorphism~\cite{rytz12,ecoop13} with no direct parallel in traditional approaches to polymorphism.  Extending  our approach for these seems sensible and feasible.

Finally, we have not considered concurrency and sequential effects, beyond noting the gap between joinoids' fork-join style operator and common source-level concurrency constructs.  
As a result we have not directly proven that our multiset-of-locks effect quantale ensures data race freedom or atomicity for a true concurrent language.

\section{Conclusions}
We have given a new algebraic characterization --- effect quantales --- for sequential effect systems, and shown it sufficient to implement complete effect systems, unlike previous approaches that focused on a subset of real language features.  We used them to model classic examples from the sequential effect system literature, and gave a syntactic soundness proof for the first generic sequential effect system.  Moreover, we give the first investigation of the generic interaction between (singleton) dependent effects and algebraic models of sequential effects, and a powerful way to derive an appropriate iteration operator on effects for many effect quantales.  We believe this is an important basis for future work designing complete sequential effect systems, and for generic effect system implementation frameworks supporting sequential effects.

\bibliographystyle{plainurl}
\bibliography{csg,effects}
\end{document}